\documentclass[11pt,a4paper]{article}
\pdfoutput=1
\usepackage{jheppub,multirow}
\usepackage{tikz}
\usepackage{color}
\usepackage{graphicx}

\newcommand{\cG}{\mathcal{G}}
\newcommand{\cN}{\mathcal{N}}
\newcommand{\cL}{\mathcal{L}}
\newcommand{\A}{{\alpha}}
\newcommand{\B}{{\beta}}
\newcommand{\D}{{\delta}}
\newcommand{\E}{{\epsilon}}

\title{
2d small N=4 Long-multiplet superconformal block}

\author{Filip~Kos$^1$,}
\author{ Jihwan~Oh$^{1,2}$}

\affiliation{
${}^1$Department of Physics,
  University of California, Berkeley, CA 94720, U.S.A.}
\affiliation{${}^2$Perimeter Institute for Theoretical Physics, 31 Caroline St. N., Waterloo, ON N2L 2Y5, Canada}
\emailAdd{filip.kos@berkeley.edu}
\emailAdd{jihwanoh@berkeley.edu}

\abstract{
We study 2d N=4 superconformal field theories, focusing on its application on numerical bootstrap study. We derive the superconformal block by utilizing the global part of the super Virasoro algebra and set up the crossing equations for the non-BPS long-multiplet 4-point function. Along the way, we build global N=4 superconformal short and long multiplets and compute all possible 2,3-point functions of long-multiplets that are needed to construct the superconformal blocks and the crossing equations. Since we consider a long-multiplet 4-point function, the number of crossing equations is huge{,} and we expect it to give a strong constraint than the usual superconformal bootstrap analysis, which relies on BPS 4-point functions. In addition, we present an alternative way to derive crossing equations using N=4 superspace and comment on a puzzle. 
}

\begin{document}
\maketitle
\flushbottom

\newcommand{\secref}[1]{\S\ref{#1}}
\newcommand{\figref}[1]{Figure~\ref{#1}}
\newcommand{\appref}[1]{Appendix~\ref{#1}}
\newcommand{\apprefrange}[2]{Appendices~\ref{#1}-\ref{#2}}
\newcommand{\tabref}[1]{Table~\ref{#1}}
\def\be{\begin{equation}}
\def\ee{\end{equation}}
\def\bear{\begin{eqnarray}}
\def\eear{\end{eqnarray}}
\def\nn{\nonumber}
\def\ie{\begin{equation}\begin{aligned}}
\def\fe{\end{aligned}\end{equation}}
\newcommand\bra[1]{{\left\langle{#1}\right\rvert}} 
\newcommand\ket[1]{{\left\lvert{#1}\right\rangle}} 

\def\Id{{\mathbf{I}}} 
\def\lra{\leftrightarrow}
\def\lea{\leftarrow}
\def\ria{\rightarrow}
\newcommand{\pa}{\partial}
\newcommand{\la}{\langle}
\newcommand{\ld}{\lambda}
\newcommand{\ra}{\rangle}
\newcommand{\cA}{\mathcal{A}}
\newcommand{\cB}{\mathcal{B}}

\newcommand{\C}{\mathbb{C}}
\newcommand{\R}{\mathbb{R}}
\newcommand{\Z}{\mathbb{Z}}
\newcommand{\CP}{\mathbb{CP}}
\newcommand{\cC}{{\mathcal C}}
\newcommand{\cD}{{\mathcal D}}
\newcommand{\cF}{{\mathcal F}}
\newcommand{\cI}{{\mathcal I}}
\newcommand{\cJ}{{\mathcal J}}
\newcommand{\cK}{{\mathcal K}}
\newcommand{\cO}{{\mathcal O}}
\newcommand{\cR}{{\mathcal R}}
\newcommand{\cS}{{\mathcal S}}
\newcommand{\cZ}{{\mathcal Z}}
\newcommand{\cW}{{\mathcal W}}
\newcommand{\cV}{{\mathcal V}}

\def\pSUSY{{\varepsilon}} 
\def\bpSUSY{{\overline{\pSUSY}}} 
\def\btheta{{\bar{\theta}}}
\def\oeta{{\overline{\eta}}}
\def\sZ{{\mathbf{Z}}} 
\def\xR{{R}}
\def\bR {\mathbb{R}}
\def\att{{\mathbf{w}}} 
\def\stt{{\mathbf{v}}} 
\def\sr{{\mathbf{r}}}
\def\pSUSY{{\varepsilon}}
\def\xTB{$\times$} 
\section{Introduction}\label{sec:Intro}
For a decade, there {has been extensive} work on solving various conformal field theories using only first principles -- unitarity, associativity of {the} operator product algebra, {and} {the} so called conformal bootstrap program. First introduced by \cite{Polyakov:1974gs,Ferrara:1973yt} and revived through \cite{Rattazzi:2008pe}, there were many attempts to solve theories with no supersymmetry \cite{Kos:2013tga,Kos:2015mba,Kos:2014bka} and different amount of supersymmetries in various dimensions \cite{Bobev:2017jhk,Bobev:2015jxa} from two dimension{s} to six dimension{s} \cite{Cornagliotto:2017dup,Lin:2016gcl,Lin:2015wcg,Chester:2014fya,Beem:2013qxa,Beem:2014zpa,Chang:2017cdx,Beem:2015aoa,Chang:2017xmr}. 

With this paper, we {wish} to fill {a} gap in the literature -- CFT in two dimension{s} with $(0,4)$ or $(4,4)$ supersymmetry \cite{Ademollo:1975an}, which seems to be the last remaining family of supersymmetric CFT{s} that {has not been explored extensively}.\footnote{\cite{Lin:2015wcg} solved {the} $K3$ $(4,4)$ SCFT, using {a} special relation between super-Virasoro block and Virasoro block available for this specific case.}At first glance, the infinite dimensional super-Virasoro symmetry \cite{Belavin:1984vu} {can} be very constraining and {provides} a lot of information by itself. However, {we are not aware of any} literature that worked out {the} super-Virasoro conformal blocks for $\cN=2$ or higher \cite{Zamolodchikov:1985ie,Belavin:2006zr,Hadasz:2006qb}, and this makes {it} difficult to use {the} full power of {the} $\cN=4$ super-Virasoro algebra in {the} bootstrap analysis.  

Still, one can try to use a global part of the superconformal algebra to construct `smaller' superconformal blocks. More precisely, we will use the fact that the 4-point correlation function of conformal primaries in the long-multiplet is decomposed into bosonic Virasoro conformal blocks, not the super-Virasoro conformal blocks. Since the coefficients {placed} in front of each decomposed Virasoro blocks are independent, the set of crossing equations is distinguished from non-supersymmetric 2d CFT 4-point functions and at the same time captures structure of $\cN=4$. This fact was used in \cite{Cornagliotto:2017dup} to do {the} $\cN=2$ long-multiplet bootstrap analysis. Our goal is to generalize this result to $\cN=4$. As we will find in {this} paper, the number of crossing equations is larger than {check} that of any numerical bootstrap literature {that we are aware of}, which makes us confident about the level of precision in using only {such} small superconformal blocks. Moreover, different from previous approaches that only analyzed particular BPS sectors of {the} theory, we set up the crossing equations using generic long-multiplets. Hence, we expect the resulting set of crossing equations to be more comprehensive{, constraining} {the} spectrum of the theory.

$2d$ $(0,4)$ or $(4,4)$ superconformal field theories {are interesting in their own right}. There are many interesting examples that {have} $\cN=4$ superconformal {symmetry} in 2-dimension{s}. Some of these include K3 $(4,4)$ theory \cite{Lin:2015wcg}, IR limits of $(0,4)$ E-string worldsheet theories \cite{Ganor:1996mu,Kim:2014dza,Witten:1997yu}, a family of $(0,4)$ theories \cite{Putrov:2015jpa} that originate from class-S theory \cite{Gaiotto:2009we,Gaiotto:2009hg}, and lastly a huge class of $(0,4)$ theories from brane-box model \cite{Hanany:2018hlz}. {Lastly, it is worth mentioning} that the 2d small $\cN=4$ chiral algebra appears in the subsector of 4d $\cN=4$ SYM \cite{Beem:2013sza,Beem:2016wfs,Bonetti:2018fqz}, which is at the same time superconformal field theory with {the} algebra $psu(2,2\rvert4)$. Although we have not attempted to study the implication of our analysis on 4d $\cN=4$ superconformal field theory, it would be very interesting to pursue {such} direction. 

Our paper is organized as follows. In \secref{sec:2dSCA}, we review {the} 2d small $\cN=4$ superconformal algebra and construct {the} supermultiplet using {the} global part of the superVirasoro algebra. In addition, we analyze short-multiplets and decompositions of long-multiplets into short-multiplets. In \secref{sec:SCBlock}, we compute superconformal blocks, starting from basic building blocks{,} such as 2-point functions and 3-point functions. A heavy amount of computation is simplified using R-symmetry and Fermion number selection rules. The solution of the system of linear equations for 3-point functions is unique and is expressed in terms of 10 independent constants that match with the counting using the superspace. This provides a strong consistency check of our calculation. With the superconformal blocks, we obtain crossing equations that can be used in the numerical analysis. In \secref{sec:superspace}, an alternative approach to compute superconformal blocks, using $\cN=4$ superspace \cite{Matsuda:1988qf,Bouwknegt:1988sv,Schoutens:1988ig}, {is} presented. We compute 3-point and 4-point invariants and construct Nilpotent invariants for superconformal block expansion {using them}. Our goal {is} to use Casimir differential equation to solve the superconformal block, but $\cN=4$ superspace does not seem to fully represent small $\cN=4$ superconformal algebra. As it is not a complete treatment, we {point} out some limitations that we encountered. We conclude the paper with future directions \secref{sec:Disc}. Since 2-point {and} 3-point function data is huge, we {include} a part of them in {Appendices} \secref{2ptnorm}, \secref{3ptnorm}, \secref{app:crossing equations}{,} and {this submission is also accompanied by} a separate Mathematica file that contains all the data.

\section{$2d$ small $\cN=4$ superconformal algebra}\label{sec:2dSCA}
In this section, we {introduce} basic elements that will be used to calculate long-multiplet n-point functions of $2d$ small $\cN=4$ global long-multiplets. In \secref{subsec:algebra}, we review $2d$ small $\cN=4$ superconformal algebra{, focusing} on the global part of the super Virasoro algebra. Following the general analysis that was done for $d\geq3$ in \cite{Cordova:2016xhm,Cordova:2016emh}, we build long- and short-multiplets in \secref{subsec:longmultiplet}, along with the decomposition of the long-multiplet into various short-multiplets. It is essential to do the short-multiplet analysis even though we {compute} long-multiplet 4-point functions, since the stress energy tensor lies in one of the short-multiplets. The identification of the multiplet that contains the stress energy tensor is crucial in the bootstrap analysis of central charges, as one needs to compute the 4-point function with stress energy tensor exchanged. Furthermore, we have identified the short-multiplets that contain the flavor current operator, which can be used in the bootstrap analysis for 2d CFT with a global symmetry.
\subsection{$\cN=4$ superconformal algebra}\label{subsec:algebra}
Let us review small $\cN=4$ superconformal algebra following \cite{Eguchi:1987sm,Eguchi:1988af,Eguchi:1987wf}. Other than the usual Virasoro algebra generators, due to enhanced supersymmetry, the superconformal algebra contains supersymmetry generators $G^a_r$, superconformal symmetry generators $\bar{G}^a_r$, and $SU(2)_R$ R-symmetry current algebra generators $T^i_m$, where $a,b$ are $SU(2)_R$ spinor indices, $i$ is $SO(3)_R$ vector index, $m\in\mathbb{Z}$, and $r\in\mathbb{Z}/2$, as we restrict ourselves in the NS sector. The super-Virasoro algebra generators satisfy {the} following (anti-)commutation relations.
\ie\nonumber
\big[L_m,L_n\big]&=(m-n)L_{m+n}+\frac{1}{2}km(m^2-1)\D_{m+n,0},\\
\{G^a_r,G^b_s\}&=\{\bar{G}^a_r,\bar{G}^b_s\}=0,\quad\{G^a_r,\bar{G}^b_s\}=2\D^{ab}L_{r+s}-2(r-s)\sigma^i_{ab}T^i_{r+s}+\frac{1}{2}k(4r^2-1)\D_{r+s,0}\D^{ab},\\
\big[T^i_m,T^j_n\big]&=i\E^{ijk}T^k_{m+n}+\frac{1}{2}km\D_{m+n,0}\D^{ij},\quad\big[T^i_m,G^a_r\big]=-\frac{1}{2}\sigma^i_{ab}G^b_{m+r},\quad\big[T^i_m,\bar{G}^a_s\big]=\frac{1}{2}\sigma^{i*}_{ab}\bar{G}^b_{m+s},\\
\big[L_m,G^a_r\big]&=(\frac{1}{2}m-r)G^a_{m+r},\quad\big[L_m,\bar{G}^a_s\big]=(\frac{1}{2}m-s)\bar{G}^a_{m+s},\quad\big[L_m,T^i_n\big]=-nT^i_{m+n}
\fe

In the following discussion, we only use the global part of the superconformal algebra to compute 2-point, 3-point, and 4-point functions. It would be far more constraining to use {the} infinite dimensional super-Virasoro algebra when one tries to bootstrap two dimensional conformal field theories, but unfortunately, the full recursion relation that leads to the approximate expression for conformal block for extended supersymmetry has not been worked out in the literature. For now, after Zamoldchikov derived the recursion relation for Virasoro conformal block \cite{Zamolodchikov:1985ie}, only $\cN=1$ super-Virasoro conformal block recursion relation was obtained \cite{Belavin:2006zr}. In spite of this limitation to use full super-Virasoro symmetry, we expect that the use of global part of super-Virasoro symmetry and Zamolodchikov recursion relation on conformal blocks should be sufficient to constrain the system and study spectrum. 

Non-trivial (anti-)commutation relations for the global part of small $\cN=4$ algebra are
\ie
&\big[L_{+1},L_{-1}\big]=2L_{0},\quad\big[L_{+1},L_{0}\big]=L_{+1},\quad\big[L_{0},L_{-1}\big]=L_{-1}\\
&\{G^a_{\pm\frac{1}{2}},G^b_{\pm\frac{1}{2}}\}=\{\bar{G}^a_{\pm\frac{1}{2}},\bar{G}^b_{\pm\frac{1}{2}}\}=0,~\{G^a_{\frac{1}{2}},\bar{G}^b_{-\frac{1}{2}}\}=2\D^{ab}L_{0}-2\sigma^i_{ab}T^i_{0},~\{G^a_{\pm\frac{1}{2}},\bar{G}^b_{\pm\frac{1}{2}}\}=2\D^{ab}L_{\pm1}\\
&\big[T^i_0,T^j_0\big]=i\E^{ijk}T^k_{0},~\big[T^i_0,G^a_{\pm\frac{1}{2}}\big]=-\frac{1}{2}\sigma^i_{ab}G^b_{\pm\frac{1}{2}},~\big[T^i_0,\bar{G}^a_{\pm\frac{1}{2}}\big]=\frac{1}{2}\sigma^{i*}_{ab}\bar{G}^b_{\pm\frac{1}{2}},\\
&\big[L_0,G^a_{\pm\frac{1}{2}}\big]=\mp\frac{1}{2}G^a_{\pm\frac{1}{2}},~\big[L_0,\bar{G}^a_{\pm\frac{1}{2}}\big]=\mp\frac{1}{2}\bar{G}^a_{\pm\frac{1}{2}},~\big[L_{\pm1},G^a_{\mp\frac{1}{2}}\big]=\pm G^a_{\pm\frac{1}{2}},~\big[L_{\pm1},\bar{G}^a_{\mp\frac{1}{2}}\big]=\pm\bar{G}^a_{\pm\frac{1}{2}}
\fe
\subsection{Long-multiplets}\label{subsec:longmultiplet}
Given the algebra, we want to construct $\cN=4$ long-multiplets labeled by superconformal primary at the bottom of the multiplets. First of all, {we} define superconformal primary operator $\cO_{h,r}$ or corresponding state $\rvert\cO_{h,r}\ra$ to be those annihilated by all positive Fourier modes of super-Virasoro algebra and eigenstates of zero modes of the algebra:
\ie
L_{n>0}\rvert\cO_{h,r}\ra=G^a_{m>0}\rvert\cO_{h,r}\ra=\bar{G}^a_{m>0}\rvert\cO_{h,j}\ra=T^i_{a>0}\rvert\cO_{h,r}\ra=0,~L_0\rvert\cO_{h,r}\ra=h\rvert\cO_{h,r}\ra,~T_0^i\rvert\cO_{h,r}\ra=(t^i)\rvert\cO_{h,r}\ra
\fe
where $h$ is holomorphic weight{,} and $r$ indicates spin $r/2$ representation of $SU(2)_R$. We will use operator $\cO_{h,r}$ and state $\rvert\cO_{h,r}\ra$ interchangeably. 

Acting $G^\A_{-\frac{1}{2}}$, $\bar{G}^\A_{-\frac{1}{2}}$ repeatedly on superconformal primary $\cO_{h,r}$ until it annihilates, one can obtain global long-multiplet $\cL_{r}$. Note that by definition of a long-multiplet, there is no null-state in the multiplet. Hence, the length of a long-multiplet is purely determined by the Fermi-statistics of raising operators. The general structure of the long-multiplet is {as follows}:
\ie
\cL_{r}=\left[\cO^{(0)}_{h,r},\bigoplus_{r_i}\cO^{(1)}_{h+\frac{1}{2},r_i},\bigoplus_{r_i}\cO^{(2)}_{h+1,r_i},\bigoplus_{r_i}\cO^{(3)}_{h+\frac{3}{2},r_i},\cO^{(4)}_{h+2,r_i}\right]
\fe
Here the superscript $(n)$ on each component indicates the number that $G$ or $\bar{G}$ act{s} on $\cO_{h,r}$; we will call half of this number as level $k=n/2$. As $G^\A_{-1/2}$ and $\bar{G}^\A_{-1/2}$ are fermionic generators, they annihilate any states after acting twice, hence the level of highest component in the long-multiplet is $4/2=2$. 

To make sure all the components $\cO^{(n)}_{h,r}$ of the long-multiplet to be (quasi)conformal primaries,  one should modify them properly, checking the (quasi)conformal primary condition: $L_{+1}\rvert\cO^{(n)}_{h,r}\ra=0$. To illustrate this point clearly, let us explicitly work out the long-multiplet built on $\phi_{h,0}$, calling it $\cL_{0}${:} 

Following diagram shows how to act $G_{-1/2}$ {and} $\bar{G}_{-1/2}$ until they annihilate superconformal primary and complete the multiplet.
\begin{center}
\includegraphics[scale=0.4]{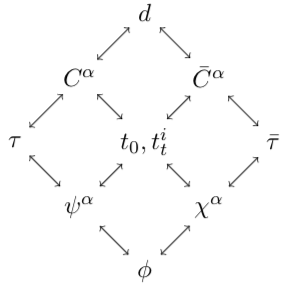}
\end{center}
Here, $G^\A_{-\frac{1}{2}}$ acts along $\nwarrow,\swarrow$ directions{,} and $\bar{G}^\A_{-\frac{1}{2}}$ acts along $\nearrow,\searrow$ directions.

In other words, each of component operator can be expressed as
\ie
&\psi^\A=G^\A\phi,~\chi^\A=\bar{G}^\A\phi,~\tau=G^\A G^\B\phi\E_{\A\B},~\bar{\tau}=\bar{G}^\A\bar{G}^\B\phi\E_{\A\B},~t_0=G^\A\bar{G}^\B\phi\E_{\A\B},~t_t^i=G^\A(\sigma^i)_{\A\B}\bar{G}^\B\phi,\\
&C^\A=G^\A G^\B\bar{G}^\gamma\phi\E_{\B\gamma},~\bar{C}^\A=G^\A\bar{G}^\B\bar{G}^\gamma\phi\E_{\B\gamma},~d=G^\A\bar{G}^\B G^\gamma\bar{G}^\D\phi\E_{\A\B}\E_{\gamma\D}
\fe
Note that we present the action of the $G^\A_{-\frac{1}{2}}$ and $\bar{G}^\A_{-\frac{1}{2}}$ as an ordered action on the state $\rvert\phi\ra$ in radial quantization using state/operator correspondence. We will stick to this convention throughout this paper. Some of the operators in the diagram do not satisfy the (quasi)conformal primary condition: $L_{+1}\rvert\cO^{(n)}_{h,r}\ra=0$, hence one needs to correct the definition of those operators so that they become conformal primaries for later use. Below, we only present the operators that are modified{, while other} operators remain same.
\ie
&\D t_0[0]=+2L[-1]\cdot\phi[0],~\D C[0]=-\frac{2h+2}{2h+1}L_{-1}\psi[0],~\D\bar{C}[0]=+\frac{2h+2}{2h+1}L_{-1}\chi[0],\\
&\D d[0]=-L_{-1}t_0[0]+\frac{2h+2}{2h+1}L_{-1}L_{-1}\phi[0]
\fe
Here $\cO[0]$ means that it is the bottom component of $SU(2)_R$ multiplet $\cO$, if $\cO$ is charged under $SU(2)_R$; otherwise, $\cO[0]$ simply means $\cO$. Other components of $SU(2)_R$ multiplet can be completed by successively acting $SU(2)_R$ raising operator $T_{-1}$ with a proper normalization. For instance,
\ie
\cO_r[n]=\frac{1}{r-(n-1)}T_{-1}\cO_r[n-1]
\fe
Similarly, we write down $\cL_{r}$ for higher $r$. For $r=1$, we have 
\begin{center}\label{generalpic}
\includegraphics[scale=0.4]{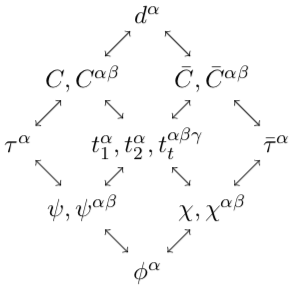}
\end{center}
In other words, each of component operator can be expressed as
\ie
&\phi^\A,~\psi^{\A\B}=G^\A\phi^\B,~\psi=G^\A\phi^\B\E_{\A\B},~\chi^{\A\B}=\bar{G}^\A\phi^\B,~\chi=\bar{G}^\A\phi^\B\E_{\A\B},\\
&\tau^\A=G^\B G^\gamma\E_{\B\gamma}\phi^\A,~\bar{\tau}^\A=\bar{G}^\B\bar{G}^\gamma\E_{\B\gamma}\phi^\A,~t_1^\A=G^\B\bar{G}^\gamma\E_{\B\gamma}\phi^\A,~t_2^\A=G^{(1}G^{2)}\phi^\A,~t_t^{\A\B\gamma}=G^\A\bar{G}^\B\phi^\gamma,\\
&C^{\A\B}=\bar{G}^\A G^\gamma\bar{G}^\D\phi^\B\E_{\gamma\delta},~C=\bar{G}^\A G^\gamma G^\delta\phi^\B\E_{\A\B}\E_{\gamma\D},~\bar{C}^{\A\B}=G^\A\bar{G}^\gamma\bar{G}^\D\phi^\B\E_{\gamma\D},~\bar{C}=G^\A\bar{G}^\gamma\bar{G}^\D\phi^\B\E_{\A\B}\E_{\gamma\D}\\
&d^\A=G^\B G^\D\bar{G}^\sigma\bar{G}^\rho\E_{\B\D}\E_{\sigma\rho}\phi^\A
\fe

Next, (quasi)conformal primary condition should be imposed. For simplicity, we will drop $[0]$ assuming all the fields shown below are bottom components of $SU(2)_R$ multiplets. 
\ie
&\D t_1^\A=\frac{2h+3}{2h}L_{-1}\phi^\A,~\D t_2^\A=\frac{-2h+3}{2h}L_{-1}\phi^\A,~\D C^{\A\B}=-\frac{2h+3}{2h+1}L_{-1}\psi^{\A\B},\\
&\D C=-\frac{2h-1}{2h+1}L_{-1}\psi,~\D\bar{C}^{\A\B}=-\frac{2h+3}{2h+1}L_{-1}\chi^{\A\B},~\D\bar{C}=+\frac{2h-1}{2h+1}L_{-1}\chi\\
&\D d^\A=-\frac{2h+1}{2(h+1)}L_{-1}t_1^\A+\frac{2h+3}{2(h+1)}L_{-1}t_2^\A
\fe
Other components of $SU(2)_R$ multiplet can be completed by successively acting $SU(2)_R$ raising operator as before.

Now, let us write down the most general $\cL_{r}$; here we include all the corrections{,} so all the operators below are (quasi)conformal primaries. To clearly illustrate $SU(2)_R$ tensor selection rules, we adopt a new convention for each operator $\cF[r][n]$, where $\cF$ is name of an operator(e.g. $\phi,\psi,\ldots$), $r$ represents the rank of $SU(2)_R$ representation, and $n$ indicates the component of $SU(2)_R$ multiplet. We denote $n=0$ as its bottom component, as before. 
\begin{center}\label{diamondgeneral}
\includegraphics[scale=0.4]{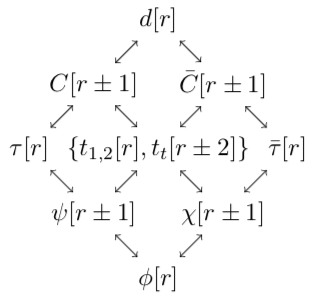}
\end{center}
Rather than writing down the whole $SU(2)_R$ multiplet, we will only present the $SU(2)_R$ bottom component of each operator for simplicity. Finally, note that $G^\A$ and $\bar{G}^\A$ are both $r=1$; we {omit} subscript $-1/2$ on these generators. 
\ie
&\psi[r-1][0]=G^1\phi[r][1]-G^2\phi[r][0],\quad\psi[r+1][0]=G^1\phi[r][0],\\
&\chi[r-1][0]=\bar{G}^1\phi[r][1]-\bar{G}^2\phi[r][0],\quad\chi[r+1][0]=\bar{G}^1\phi[r][0],\\
&\tau[r][0]=G^1G^2\phi[r][0],\quad\bar{\tau}[r][0]=\bar{G}^1\bar{G}^2\phi[r][0],\\
&t_t[r-2][0]=G^1\bar{G}^1\phi[r][2]-G^1\bar{G}^2\phi[r][1]-G^2\bar{G}^1\phi[r][1]+G^2\bar{G}^2\phi[r][0],\\
&t_1[r][0]=G^1\bar{G}^1\phi[r][1]-G^1\bar{G}^2\phi[r][0]+\frac{2h+r+2}{2h}L_{-1}\phi[r][0],\\
&t_2[r][0]=G^1\bar{G}^1\phi[r][1]-G^2\bar{G}^1\phi[r][0]+\frac{-2h+r+2}{2h}L_{-1}\phi[r][0],\\
&t_t[r+2][0]=G^1\bar{G}^1\phi[r][0],\\
&C[r-1][0]=\bar{G}^1G^1G^2\phi[r][1]-\bar{G}^2G^1G^2\phi[r][0]-\frac{2h-r}{2h+1}L_{-1}\psi[r-1][0],\\
&C[r+1][0]=\bar{G}^1G^1G^2\phi[r][0]-\frac{2h+2+r}{2h+1}L_{-1}\psi[r+1][0],\\
&\bar{C}[r-1][0]=G^1\bar{G}^1\bar{G}^2\phi[r][1]-G^2\bar{G}^1\bar{G}^2\phi[r][0]+\frac{2h-r}{2h+1}L_{-1}\chi[r-1][0],\\
&\bar{C}[r+1][0]=G^1\bar{G}^1\bar{G}^2\phi[r][0]+\frac{2h+r+2}{2h+1}L_{-1}\chi[r+1][0],\\
&d[r][0]=G^1G^2\bar{G}^1\bar{G}^2\phi[r][0]+\sum_{i=1}^2(-1)^i\frac{2+2h+(-1)^ir}{2(h+1)}L_{-1}t_i[r][0]+\frac{4h+4h^2-2r-r^2}{2h(1+2h)}L_{-1}L_{-1}\phi[r][0]
\fe
\subsection{Short-multiplets}\label{shortm}
Superconformal algebra determines a shortening condition for the long-multiplet. General analysis was done in \cite{Cordova:2016xhm,Cordova:2016emh} for higher dimension $3\leq d\leq6$. We will use their insights to analyze our case and sometimes adopt their conventions. 

Recall
\ie
\{G^a_{\frac{1}{2}},\bar{G}^b_{-\frac{1}{2}}\}=2\D^{ab}L_{0}-2\sigma^{ab}_iT^i_{0}
\fe
By sandwiching between two superconformal primary states $\rvert\phi_{h,r}\ra$ and imposing unitarity, one gets
\ie
\la\phi_{h,r}\rvert\{G^a_{\frac{1}{2}},\bar{G}^b_{-\frac{1}{2}}\}\rvert\phi_{h,r}\ra=\la\phi_{h,r}\rvert2\D^{ab}L_{0}-2\sigma_i^{ab}T^i_{0}\rvert\phi_{h,r}\ra=2(h-\frac{r}{2})\geq0
\fe
This implies that the multiplet is shortened when the superconformal primary satisfies {the} $h=\frac{r}{2}$ condition. By looking at the algebra, one can easily see that only this specific type of the anti-commutator gives the non-trivial shortening condition that gives zero in the norm of descendants, as it is clear in the explicit calculation given in Appendix \secref{2ptnorm}.

Let us apply this to $\cL_0$, $\cL_1$, {and $\cL_{r\geq2}$} separately. For $\cL_0$, as $h[\phi]\ria0$, only  superconformal primary that survives is
\ie
\phi={\bf{1}}
\fe
This is the unit operator of CFT. Let us denote it as $\cA_0$.

For $\cL_1$, as $h[\phi^\A]\ria\frac{1}{2}$, there is one short-multiplet, as shown below. $\phi^\A$ is a two-component fermion{,} and $\psi,\chi$ are bosons. We denote it as $\cA_{1}$.
\begin{center}
\includegraphics[scale=0.4]{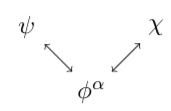}
\end{center}
This multiplet should be the one that contains a flavor current operator. The reason is {the} following. As a flavor symmetry commutes with the superconformal symmetry, the superconformal multiplet that contains the flavor current operator should place it at the top of the multiplet. One can see the top component of this short-multiplet does not carry $SU(2)_R$ index, consistent with the flavor symmetry current operator being R-symmetry neutral. Furthermore, we know that the conformal weight of $\{\psi,\chi\}$ is 1, which is the correct dimension of flavor current operator. Also, flavor symmetry current operator can not reside in the long-multiplet, as the top-component of any long-multiplet should have conformal weight 2, at least.

For $\cL_{r\geq2}$, as $h[\phi[r\geq2]]\ria\frac{r}{2}$, there is one short-multiplet that appears at the bottom corner. We denote it as $\cA_r$.
\begin{center}\label{small}
\includegraphics[scale=0.4]{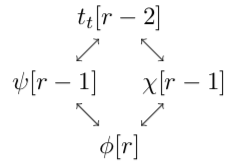}
\end{center}
For $r=2$, it is natural to think that the holomorphic stress-energy tensor lives in this short-multiplet as a top component. First, the top component has {the} desired quantum number: $(h,r)=(2,0)$. Second, as the stress energy tensor should commute with global super(conformal)symmetry generators $G^\A_{-\frac{1}{2}}$ {and} $\bar{G}^\A_{-\frac{1}{2}}$, it should be on top of multiplet. Furthermore, other components of the multiplet reproduce the desired content of the stress-energy multiplet: $SU(2)_R$ R-symmetry current operator with $SU(2)_R$ rank-2 at the bottom and the global super(conformal) currents with $SU(2)_R$ rank-1 in the middle. Of course, each operator in the multiplet has expected conformal dimension: $1$, $\frac{3}{2}$, {and} $2$. 

In $\cN=2$ superconformal field theory, a stress energy tensor lives in $\cN=2$ long-multiplet \cite{Cornagliotto:2017dup}. The $\cN=2$ long-multiplet is a short-multiplet in the point of view of $\cN=4$ theory. Above analysis shows that in $\cN=4$ theory, the stress energy tensor should live in the short-multiplet, different from $\cN=2$ case.
\subsection{Decomposition of the Long-multiplets into the Short-multiplets}\label{decomp}
Similar to that of higher dimensional superconformal field theories, 2d $\cN=4$ long-multiplet has a decomposition into the short-multiplets. We could see all $2d$ $\cN=4$ short-multiplets that appear in the decomposition of the long-multiplet are `Short-multiplet at Threshold', in the terminology of \cite{Cordova:2016emh}.

Let us illustrate this point with $\cL_0$, $\cL_1$ long-mutiplets.
\begin{center}
\includegraphics[scale=0.4]{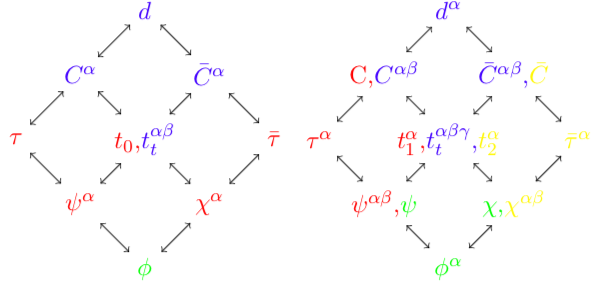}
\end{center}
The green threshold short-multiplets of $\cL_0$, $\cL_1$ first decouple, as $h[\phi[0]]\ria0$ {and} $h[\phi[1]]\ria\frac{1}{2}$. Moreover, red and yellow operators form $\cL_1$, $\cL_2$ short-multiplets that were called as $\cA_1$, $\cA_2$. Finally, it can be checked that the top blue short-multiplets are the short-multiplets of $\cL_2$, $\cL_3$ at threshold. Hence, the long-multiplets $\cL_0$, $\cL_1$ decompose when superconformal primaries {saturate} the unitarity bound as
\ie
\lim_{h\ria0}\cL_0[h]&\ria\cA_0[0]\oplus\cA_1[\frac{1}{2}]\oplus\bar{\cA}_1[\frac{1}{2}]\oplus\cA_2[1]\\
\lim_{h\ria\frac{1}{2}}\cL_1[h]&\ria\cA_1[\frac{1}{2}]\oplus\cA_2[1]\oplus\bar{\cA}_2[1]\oplus\cA_3[\frac{3}{2}]
\fe
where we used convention $\cF_r[h]$ for long or short multiplet with rank-r $su(2)_R$ representation and conformal weight $h$.

More generally,
\ie\label{short}
\lim_{h\ria\frac{r}{2}}\cL_r[h]\ria\color{green}{\cA}_r[\frac{r}{2}]\oplus\color{red}\cA_{r+1}[\frac{r+1}{2}]\oplus\color{blue}\bar{\cA}_{r+1}[\frac{r+1}{2}]\oplus\color{purple}\cA_{r+2}[\frac{r+2}{2}]
\fe
\begin{center}
\includegraphics[scale=0.4]{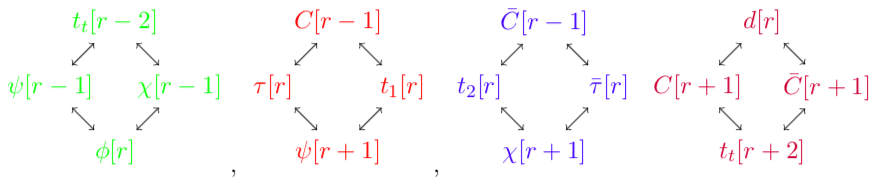}
\end{center}
From this, we can see {the shortening condition in 2d $\cN=4$ superconformal algebra,} and the kind of short-multiplet that could appear is simpler compared to higher dimension analogue \cite{Cordova:2016emh}. This is not surprising as there is no non-trivial Lorentz symmetry index, unless combined with Left-moving non-SUSY side, and the R-symmetry algebra is simple in $2d$ superconformal field theory.

The short-multiplet structures can also be read off from the direct calculation of two-point function. We sketch the calculation in the next section and present the results in the appendix \secref{2ptnorm}. In short, two-point functions constructed from $\cL_0$, $\cL_1$, and $\cL_2$ have zero at $h=0$, $h=\frac{1}{2}$, $h=1$, respectively. They are unique zeros for each multiplet and the highest degree is 2, as the $G$, $\bar{G}$ anti-commutator can at most appear twice when we build a long-multiplet, due to the grassmann nature of the supersymmetry generators.
\section{Superconformal block computation}\label{sec:SCBlock}
The main object to study is 4-point function of identical rank-0 long-multiplet $\cL_0$. From now, we will interchangeably use $\cL_0$ and $\Phi_i(Z_i)$ to denote rank-0 long-multiplet. In superspace, $\Phi_i(Z_i)$ has {the} following expansion with proper $SU(2)_R$ index contraction assumed:
\ie
\Phi(Z)=\phi(z)+\psi\bar{\theta}+\chi\theta+\tau\theta\theta+\bar{\tau}\bar{\theta}\bar{\theta}+t_0\theta\bar{\theta}+t_t\theta\sigma\bar{\theta}+C\theta\bar{\theta}\bar{\theta}+\bar{C}\theta\theta\bar{\theta}+d\theta\theta\bar{\theta}\bar{\theta}
\fe

One way to study 4-point function is to work in the superspace, as it provides a natural framework to use the superconformal algebra to fix the structure of 4-point function and selection rules to classify non-trivial component 4-point functions, such as $\la\phi_1\phi_2\phi_3\phi_4\ra$, $\la\psi_1\chi_2\phi_3\phi_4\ra$, {and} $\la\tau_1\phi_2\phi_3\phi_4\ra$. However, we found $\cN=4$ superspace has a subtlety that prevented us to use it to compute the superconformal blocks. Still, we could proceed to compute component 4-point functions by classical method in computing n-point function and superconformal algebra that we will describe in this section. We will separately discuss $\cN=4$ superspace in the next section, up to the point that we could reach and comment on the subtle point.

We will compute all possible 4-point functions of component operators in $\cL_0$: 
\ie
\{\phi,\psi^\A,\chi^\A,\tau,\bar{\tau},t_0,t_t,C^\A,\bar{C}^\A,d\}
\fe 
If we treat different $SU(2)_R$ index $\A=1,2$ separately, in principle there are $16^4$ possible 4-point functions to compute. The number grows tremendously if we include $\la\cL_0\cL_0\cL_r\ra$ three point functions. Of course, Fermion number and $SU(2)_R$ symmetry selection rules help to restrict the set to a reasonably small subset. 

Let us start with the simplest one $\la\phi_1\phi_2\phi_3\phi_4\ra$ to illustrate the strategy to get {the} conformal block decomposition of general 4-point functions. Here, $\phi_i$ are identical superconformal primaries of long-multiplet $\cL_0$. Note that although we used different indices to distinguish their positions in the superspace for the operators $\phi_i$, they are essentially identical superconformal primaries with same $h$:
\ie\label{4ptf}
\la\phi_1(x_1)\phi_2(x_2)\phi_3(x_3)\phi_4(x_4)\ra&=\sum_{\cO,\cO'}f_{\phi_1\phi_2\cO}f_{\phi_3\phi_4\cO'}C_a(x_{12},\pa_2)C_b(x_{34},\pa_4)\la\cO^a(x_2){\cO'}^b(x_4)\ra\\
&=\sum_\cO f_{\phi\phi\cO}f_{\phi\phi\cO'}C_a(x_{12},\pa_2)C_b(x_{34},\pa_4)\frac{f_{\cO\cO'}}{x^{2h_\cO}_{24}}\\
&=\frac{1}{x^{2h_\phi}_{12}x^{2h_\phi}_{34}}\sum_\cO \frac{f_{\phi\phi\cO}f_{\phi\phi\cO'}}{f_{\cO\cO}}g_{h_\cO}(z)
\fe
where $z$ is the standard bosonic cross-ratio $z=\frac{x_{12}x_{34}}{x_{13}x_{24}}$, and we used
\ie
&\la\cO(x_1){\cO'}(x_2)\ra=\frac{f_{\cO\cO'}}{x_{12}^{2h_\cO}}\\
&\la\phi_1(x_1)\phi_2(x_2)\cO(x_0)\ra=\frac{f_{\phi_1\phi_2\cO}}{x_{12}^{h_1+h_2-h_0}x_{20}^{h_2+h_0-h_1}x_{10}^{h_1+h_0-h_2}}\\
&g_{h}(z)\equiv x^{h}_{12}x^{h}_{34}C_a(x_{12},\pa_2)C_b(x_{34},\pa_4)\frac{f^2_{\cO\cO'}}{x^{2h}_{24}}
\fe
Here $x_{ij}=x_i-x_j$ and $x_i$ are holomorphic coordinates. \footnote{The factorization of correlation functions does not hold in general. So, by this we assume that the factorization holds for our case. We thank an anonymous referee of JHEP, who pointed out this subtlety that we were not aware of.} 

We can decompose each of 4-point functions into bosonic blocks:
\ie\label{bosonicblock}
\la\phi_1(x_1)\phi_2(x_2)\phi_3(x_3)\phi_4(x_4)\ra&=\frac{1}{x^{2h_\phi}_{12}x^{2h_\phi}_{34}}\bigg(\sum_{\{\cO_1\}} \frac{f_{\phi\phi\cO_1}f_{\phi\phi\cO'_1}}{f_{\cO_1\cO'_1}}g_{h_{\cO_1}}(z)+\ldots+\sum_{\{\cO_5\}}\frac{f_{\phi\phi\cO_5}f_{\phi\phi\cO'_5}}{f_{\cO_5\cO'_5}}g_{h_{\cO_5}}(z)\bigg)
\fe
where $\cO_i$ represents a conformal primary in the long-multiplet $\cL_r$ that appears in the OPE of $\phi$ and $\phi$. Here, $i$ is $2k+1$, where $k$ is the level of the conformal primary in $\cL_r$. This decomposition is the essential property for the long-multiplet 4-point function analysis, since it provides a detour from the use of the unknown $\cN=4$ super-Virasoro conformal blocks. One might think that the coefficients in front of $g_{h_{\cO_i}}$ may be dependent, but this is not the case as can be seen in the explicit computation of 3-point functions shown in the subsequent sub-sections. The independence of the coefficients in the 4-point function decomposition indicates the novelty of our $\cN=4$ study, distinguished from the bootstrap of non-supersymmetric 2d CFT.

Due to Zamolodchikov \cite{Zamolodchikov:1985ie}, approximate expression for $g_h(z)$ is known and it can be recursively deduced from $sl(2)$ bosonic conformal block
\ie
g^{h_{12},h_{34}}_{h}(z)=z^h{}_2F_1(h-h_{12},h+h_{34},2h,z)
\fe
Hence, what remains to compute is 3-point function coefficients $f_{\phi\phi\cO_n}$ and 2-point function normalization $f_{\cO_n\cO'_n}$. 

Similarly, it is easy to generalize to any component 4-point functions, $\la p_1p_2p_3p_4\ra$. In general,
\ie\label{scb}
\la p_1p_2p_3p_4\ra&=\frac{1}{x^{2h_{p_1}}_{12}x^{2h_{p_3}}_{34}}\bigg(\sum_{\{\cO_1\}}\frac{f_{p_1p_2\cO_1}f_{p_3p_4\cO'_1}}{f_{\cO_1\cO'_1}}g_{h_{\cO_1}}(z)+\ldots+\sum_{\{\cO_i\}}\frac{f_{p_1p_2\cO_n}f_{p_3p_4\cO'_n}}{f_{\cO_n\cO'_n}}g_{h_{\cO_n}}(z)\bigg)
\fe
Note that the exchange operators $\cO_i,\cO'_i$ can belong to any rank-r supermultiplet $\cL_r$, not just $\cL_0$ where all 4 external operators belong to. We can classify blocks shown in \eqref{scb} in terms of what super-multiplet $\{\cO_i\}$ belongs to. There are three possible supermultiplets that participate in \eqref{scb}; they are $\cL_0$, $\cL_1$, {and} $\cL_2$. As before, the necessary computation reduces {to figuring} out non-trivial $f_{p_1p_2\cO}$, $f_{\cO'p_3p_4}$ and $f_{\cO\cO'}$. 
\subsection{Selection rules}
There are two selection rules that we will use frequently in the subsequent sections{:} 1. Fermion number selection rule, 2. R-symmetry selection rule. For the first selection rule, we assign Fermion number to each operator in $\cL_0$, $\cL_1$, {and} $\cL_2$:
\begin{center}
\begin{tabular}{|c|c|c|c|c|c|c|c|c|c|c|} \hline
$\cL_0$  &  $\phi$&  $\psi^\A$ &$\chi^\A$ & $\tau$ & $\bar{\tau}$& $t_0$& $t^i_t$ & $C^\A$ & $\bar{C}^\A$ & d\\
\hline
            F &        0 &               1 &        1     &     0      &        0         &     0    &      0    &      1      &       1             &0\\
\hline
\end{tabular}
\begin{tabular}{|c|c|c|c|c|c|c|c|c|c|c|c|c|c|c|c|} \hline
$\cL_1$  &  $\phi^\A$&  $\psi$ & $\psi^{\A\B}$ & $\chi$ & $\chi^{\A\B}$ & $\tau^\A$ & $\bar{\tau}^\A$& $t^\A_1$& $t^\A_2$& $t^{\A\B\gamma}_t$ & $C$ & $C^{\A\B}$ & $\bar{C}$ & $\bar{C}^{\A\B}$ & $d^\A$\\
\hline
            F &        1 & 0              &        0             &     0      &        0         &     1    &      1    &      1      &       1         & 1    &0&0&0&0&1\\
\hline
\end{tabular}
\end{center}
\begin{center}
\begin{tabular}{|c|c|c|c|c|c|c|c|c|c|c|c|c|c|c|c|c|} \hline
$\cL_2$  &  $\phi^{\A\B}$&  $\psi^\A$ & $\psi^{\A\B\gamma}$ & $\chi^\A$ & $\chi^{\A\B\gamma}$ & $\tau^{\A\B}$ & $\bar{\tau}^{\A\B}$& $t$& $t_1^{\A\B}$& $t_2^{\A\B}$ & $t^{\A\B\gamma\D}$& $C^\A$ & $C^{\A\B\gamma}$ & $\bar{C}^\A$ & $\bar{C}^{\A\B\gamma}$ & $d^{\A\B}$\\
\hline
            F &        0 & 1              &     1             &     1      &        1         &     0    &      0    &      0      &       0         & 0 &0    &1&1&1&1&0\\
\hline
\end{tabular}
\end{center}
For n-point function $\la\cF_1\ldots\cF_n\ra$ not to vanish, the sum of fermion number should be even.
\ie
\sum_{i=1}^nF[\cF_i]=0~\text{mod}~2
\fe

Next, in describing the R-symmetry selection rules, we will take the general notations that were used in illustrating the primary operators in the general $\cL_r$ multiplet. There are two rules: $U(1)_R\subset SU(2)_R$ charge conservation rule and $SU(2)_R$ selection rule. For 3-point function $\la\cF_1[r_1][n_1]\cF_2[r_2][n_2]\cF_3[r_3][n_3]\ra$, the first rule is 
\ie
\frac{r_1+r_2+r_3}{2}-(n_1+n_2+n_3)=0
\fe
and the second rule is
\ie
\rvert r_1-r_2\rvert\leq r_3 \leq r_1+r_2
\fe
For 4-point function $\la\cF_1[r_1][n_1]\cF_2[r_2][n_2]\cF_3[r_3][n_3]\cF_4[r_4][n_4]\ra$, the first rule is
\ie
\frac{r_1+r_2+r_3+r_4}{2}-(n_1+n_2+n_3+n_4)=0
\fe
and the second rule is
\ie
\min[r_1+r_2,r_3+r_4]\geq\max[\rvert r_1-r_2\rvert,\rvert r_3-r_4\rvert]
\fe
\subsection{2-point functions}\label{subsec:2pt}
Let us start with the simplest case: 2-point function normalization $f_{\cO\cO'}$. A simple fact that super(conformal) symmetry generator annihilates vacuum leads to {the} following equation
\ie
\la0\rvert\cF_1\cF_2G^\A\rvert0\ra=0,\quad\la0\rvert\cF_1\cF_2\bar{G}^\A\rvert0\ra=0
\fe
where $\cF_i\in\{\phi,\psi,\tau,\tau,\bar{\tau},t_0,t_1,t_2,t_t,C,\bar{C},d\}$ are component primary operators in supermultiplets $\cL_0$, $\cL_1$, {and} $\cL_2$.

Commuting $G^\A$, $\bar{G}^\A$ to the left generates a set of linear equations for each pair $\{\cF_1,\cF_2\}$.
\ie\label{2pteqn}
&(-1)^{f_2}\la\cF_1[G^\A,\cF_2]\ra+(-1)^{f_1+f_2}\la[G^\A,\cF_1]\cF_2\ra=0\\
&(-1)^{f_2}\la\cF_1[\bar{G}^\A,\cF_2]\ra+(-1)^{f_1+f_2}\la[\bar{G}^\A,\cF_1]\cF_2\ra=0,
\fe
where factors of $(-1)$ is due to the commutation of $\cF_i$ and fermionic operator $G^\A$. $f_i$ is 0 for $\cF_i$ boson, and 1 for $\cF_i$ fermion. Here, $G^\A\cF_i$ {or} $\bar{G}^\A\cF_i$ can be computed by utilizing the superconformal algebra and is equal to a linear combination of $\cF_j$ with known proportionality constants $c^\A_{j,k}$, $\bar{c}^\A_{j,k}$ with possible corrections $L_{-1}\cF_i$, $L_{-1}L_{-1}\cF_i$, which we have seen in the definitions of conformal primaries {above}.
\ie
G^\A\cF_i=\sum_{k=1}^{n_i}c^\A_{i,k}\cF_k,\quad\bar{G}^\A\cF_i=\sum_{k=1}^{n_i}\bar{c}^\A_{i,k}\cF_k
\fe
As 2-point functions are fixed up to normalization constants
\ie\label{2pt}
\la\cF_i\cF_j\ra=\frac{f_{\cF_i\cF_j}}{(z_1-z_2)^{2h_i}}{,}
\fe
the above equations \eqref{2pteqn} become
\ie\label{222}
&(-1)^{f_2}\sum_{i=1}^{n_2}\frac{c_{2,i}f_{\cF_1\cF_i}}{(z_1-z_2)^{2h_1}}+(-1)^{f_1+f_2}\sum_{i=1}^{n_1}\frac{c_{1,i}f_{\cF_i\cF_2}}{(z_1-z_2)^{2h_2}}=0\\
&(-1)^{f_2}\sum_{i=1}^{n_2}\frac{\bar{c}_{2,i}f_{\cF_1\cF_i}}{(z_1-z_2)^{2h_1}}+(-1)^{f_1+f_2}\sum_{i=1}^{n_1}\frac{\bar{c}_{1,i}f_{\cF_i\cF_2}}{(z_1-z_2)^{2h_2}}=0
\fe
By factoring out the common denominators, \eqref{222} becomes
\ie
&(-1)^{f_2}\sum_{i=1}^{n_2}c_{2,i}f_{\cF_1\cF_i}(z_1-z_2)^{2h_2}+(-1)^{f_1+f_2}\sum_{i=1}^{n_1}c_{1,i}f_{\cF_i\cF_2}(z_1-z_2)^{2h_1}=0\\
&(-1)^{f_2}\sum_{i=1}^{n_2}\bar{c}_{2,i}f_{\cF_1\cF_i}(z_1-z_2)^{2h_2}+(-1)^{f_1+f_2}\sum_{i=1}^{n_1}\bar{c}_{1,i}f_{\cF_i\cF_2}(z_1-z_2)^{2h_1}=0
\fe
from which one can read off the coefficients of $z_1^Iz_2^J$ that is a linear system of $f_{\cF_1\cF_i}$, $f_{\cF_i\cF_2}$. We can easily solve the linear system to fix all $f_{\cF_i\cF_j}$ up to three independent constants. Each of the three constants comes from $\cL_0$, $\cL_1$, {and} $\cL_2$, respectively. The three constants can be fixed to 1 in the later computation. 

We should obtain all non-trivial 2-point functions of $\cL_0$, $\cL_1$, {and} $\cL_2$. The reason that we do not consider higher rank $\cL_r$ with $r>2$ will become clear in the next subsection \secref{subsec:3pt} where we discuss 3-point function. In practical computation, because of {the} large number of operators in $\cL_0$, $\cL_1$, {and} $\cL_2$, it would be better to first restrict the set of non-trivial two-point functions by using {the} 2-point function definition and $SU(2)_R$ symmetry selection rules. \\For instance, \\1. $\la\phi\phi G^\A\ra=\la\phi\psi^\A\ra+\la\psi^\A\phi\ra$ will not give any non-trivial condition as both $\la\phi\psi^\A\ra$ and $\la\psi^\A\phi\ra$ vanish since $\phi$ and $\psi^\A$ have different conformal weight. \\2. Equations from $\la\phi\psi^1G^1\ra$ are trivial, since they are equal to $\la\phi0\ra+\la\psi^1\psi^1\ra$, and $\la\psi^1\psi^1\ra$ vanishes due to the $SU(2)_R$ selection rules.

It would be instructive to explicitly work out one non-trivial example that passed the two simple tests above, as we will use this procedure to construct higher n-point functions. Start from $\la\chi^2\bar{\tau}G^1\ra$, where both $\chi$ and $\bar{\tau}$ are in $\cL_0$ built from conformal weight $h$ superconformal primary $\phi$.
\ie\label{ex1}
0=\la\chi^2\bar{\tau}G^1\ra&=\la\chi^2[G^1,\bar{\tau}]\ra-\la [G^1,\chi^2]\bar{\tau}\ra
\fe
From the superconformal algebra, we know
\ie
G^1\bar{\tau}=-\bar{C}^2+\frac{2+2h}{1+2h}\pa\chi^2,\quad G^1\chi^2=-t^3_t
\fe
So, \eqref{ex1} becomes
\ie
0=\la\chi^2\bar{\tau}G^1\ra&=-\la\chi^2\bar{C}^2\ra+\frac{2+2h}{1+2h}\pa_{z_2}\la\chi^2\chi^2\ra+\la t_t^3\bar{\tau}\ra
\fe
$\la\chi^2\bar{C}^2\ra$ vanishes due to the R-symmetry selection rules. For {the} next two terms, we substitute explicit 2-point function formula \eqref{2pt} and get
\ie
0&=0+\frac{2+2h}{1+2h}\pa_{z_2}\frac{f_{\chi^2\chi^2}}{(z_1-z_2)^{1+2h}}+\frac{f_{t_t^3\bar{\tau}}}{(z_1-z_2)^{2+2h}}\\
0&=(2+2h)f_{\chi^2\chi^2}+f_{t_t^3\bar{\tau}}
\fe
So, we obtained one linear equation that relates two 2-point function normalizations. Similarly, we can do the same thing for $\la\chi^2\bar{\tau}G^2\ra$, $\la\chi^2\bar{\tau}\bar{G}^1\ra$, {and} $\la\chi^2\bar{\tau}\bar{G}^2\ra$. We automated this procedure in Mathematica to compute all non-trivial 2-point function normalizations $f_{\cF_i\cF_j}$. For simplicity, let us only present those of $\cL_0$. They are fixed up to one constant denoted as $f_{\phi\phi}$. Of course, most of them vanish by the definition of {a} 2-point function.
\ie
&f_{\phi\phi},~f_{\psi^1\chi^2}=f_{\chi^2\psi^1}=-2hf_{\phi\phi},~f_{\psi^2\chi^1}=f_{\chi^1\psi^2}=-2hf_{\phi\phi},~f_{\tau\bar{\tau}}=f_{\bar{\tau}\tau}=-4h(1+h)f_{\phi\phi},\\
&f_{t_0t_0}=8h(1+h)f_{\phi\phi},~f_{t_t^1t_t^3}=f_{t_t^3t_t^1}=-4h^2f_{\phi\phi},~f_{t_t^2t_t^2}=2h^2f_{\phi\phi},~f_{C^1\bar{C}^2}=f_{\bar{C}^2C^1}=-\frac{16h^2(h+1)^2}{1+2h}f_{\phi\phi}\\
&f_{C^2\bar{C}^1}=-f_{\bar{C}^1C^2}=\frac{16h^2(1+h)^2}{1+2h}f_{\phi\phi},~f_{dd}=\frac{16h^2(1+h)^2(3+2h)}{1+2h}f_{\phi\phi}
\fe
Similarly, 2-point functions that consist of $\cL_1$ and $\cL_2$ are fixed up to one constant respectively.  
\subsection{3-point functions}\label{subsec:3pt}
We want to compute 3-point function OPE coefficients $f_{\cF_1\cF_2\cF_3}$, where $\cF_1,\cF_2\in\cL_0$, $\cF_3\in\cL_0,\cL_1,\cL_2,\cL_3,\cL_4,\cL_5$, as $\cF_1,\cF_2$ are two of external primary operators in the 4-point function and $\cF_3$ is an exchanged primary operator that can in principle be in any of $\cF_i$, although we will find out 3-point function with $\cF_3\in\cL_3,\cL_4,\cL_5$ vanishes.

$n-$point correlator calculation becomes extremely complicated from $n\geq3$, as the total number of possible 3 point combination increases tremendously compared to that of 2-point function case. 
\begin{center}
\begin{tabular}{|c|c|c|c|c|c|c|} \hline
 $\cF_3\in$  &  $\cL_0$&  $\cL_1$ &$\cL_2$ & $\cL_3$ & $\cL_4$& $\cL_5$ \\
\hline
$\#$ \text{of} $f_{\cF_1\cF_2\cF_3}$ & $16^3$ & $16^3\times2$ &  $16^3\times3$ &  $16^3\times4$ &  $16^3\times5$ &  $16^3\times6$\\
\hline
\end{tabular}
\end{center}
Hence, we need to introduce more systematic way of selecting non-trivial equations by refining $SU(2)_R$ selection rules. We used the rules to construct a linear system of 2-point functions, but as it starts to impose non-trivial constraints starting from 3-point function, we describe an additional procedure here.

Before deriving a system of equations from $G^\A$, $\bar{G}^\A$ commutation, it may be more efficient to use $SU(2)_R$ generators $T_0^+$ to obtain an extra set of equations that prepares a smaller subset of the entire set on which we apply $G^\A$, $\bar{G}^\A$ commutation procedure. The method is essentially the same as before with $T_0^+$ replacing $G^\A$, $\bar{G}^\A$ that leads
\ie
\la0\rvert\cF_1\cF_2\cF_3T^+_0\rvert0\ra=0
\fe
where $\cF_i\in\{\phi,\psi,\tau,\tau,\bar{\tau},t_0,t_1,t_2,t_t,C,\bar{C},d\}$ are component primary operators in the supermultiplets $\cL_0$, $\cL_1$, $\cL_2$.

By commuting $T^+_0$ to the left, we get a set of linear equations
\ie
\la\cF_1\cF_2[T^+_0,\cF_3]\ra+\la\cF_1[T^+_0,\cF_2]\cF_3\ra+\la\cF_1\cF_2[T^+_0,\cF_3]\ra=0
\fe
As $T^+_0$ is a raising operator of $SU(2)_R$ R-symmetry algebra, it is a map from one operator to the same operator with different $SU(2)_R$ index. For instance, $T^+_0:~\psi^1\ria\psi^2$. As a result, it will not be as powerful as the constraints from the equations of $G^\A$, $\bar{G}^\A$ commutation; however, it completely reduces $SU(2)_R$ degeneracy and enables us to only consider one component of each $SU(2)_R$ multiplet. Especially, this procedure helps us to reduce a significant number of degrees of freedom in higher rank representations in $\cL_1$ and $\cL_2$. Following table shows how much the number of operators in each $\cL_r$ is reduced after using $T^+_0$.
\begin{center}
\begin{tabular}{|c|c|c|c|c|c|c|} \hline
 Rank  &  0&  1 & 2 & 3 & 4 &5 \\
\hline
\text{Before} & 16 & 32 & 48 & 64 & 80 & 96\\
\hline
\text{After}    & 10 & 15 & 16 & 16 & 16 &16\\
\hline
\end{tabular}
\end{center}

Let us work out explicitly, for example, $\la\chi^1\bar{C}^2t_t^2T^+_0\ra$.
\ie\label{ex3}
0=\la\psi^2\bar{C}^2t_t^2T^+_0\ra&=\la\psi^2\bar{C}^2[T^+_0,t_t^2]\ra+\la\psi^2[T^+_0,\bar{C}^2]t_t^2\ra+\la[T^+_0,\psi^2]\bar{C}^2t_t^2\ra\\
&=\la\psi^2\bar{C}^2t_t^1\ra+\la\psi^2\bar{C}^1t_t^2\ra+\la\psi^1\bar{C}^2t_t^2\ra
\fe
As three terms in the last line of \eqref{ex3} share the same set of conformal weights, $(h+\frac{1}{2},h+\frac{3}{2},h+1)$, $z_i$ dependence is gone and the above equation becomes a linear equation of three 3-point coefficients.
\ie
f_{\psi^2\bar{C}^2t^1_t}+f_{\psi^2\bar{C}^1t^2_t}+f_{\psi^1\bar{C}^2t^2_t}=0
\fe

In this way, we can reduce the number of 3-point functions that we need to treat in $G$, $\bar{G}$ commutation equations. Following table shows the reduction of the number of the independent 3-point functions $\la\cL_0\cL_0\cL_i\ra$ after using Fermion number, R-symmetry selection rules, and $T^+_0$ commutation.
\begin{center}
\begin{tabular}{|c|c|c|c|c|c|c|} \hline
i &  0&  1 & 2 & 3 & 4 &5 \\
\hline
\text{Before} & 4096 & 8192 & 12288 & 16384 & 20480 & 24576\\
\hline
\text{Fermion Number} & 2048 & 4096 & 6144 & 8192 & 10240 & 12288\\
\hline
\text{R-symmetry} & 1364 & 1364 & 1315 & 840 & 341 & 84\\
\hline
$T^+_0$    & 429 & 572 & 429 & 208 & 64 &12\\
\hline
\end{tabular}
\end{center}

Next, let us work out one simple case from $G$, $\bar{G}$ commutations. 
\ie\label{threeptex}
0&=\la\phi\phi\chi^1 G^1\ra=-\la\phi\phi G^1\chi^1\ra-\la\phi G^1\phi\chi^1\ra-\la G^1\phi\phi\chi^1\ra\\
&=\la\phi\phi t_t^2\ra+\frac{1}{2}\la\phi\phi t_0\ra-\pa_{z_3}\la\phi\phi\phi\ra-\la\phi\psi^1\chi^1\ra-\la\psi^1\phi\chi^1\ra\\
&=\frac{f_{\phi\phi t^2_t}}{z_{12}^{h-1}z_{23}^{h+1}z_{31}^{h+1}}+\frac{1}{2}\frac{f_{\phi\phi t_0}}{z^{h-1}_{12}z^{h+1}_{23}z^{h+1}_{31}}-\pa_{z_3}\frac{f_{\phi\phi\phi}}{z_{12}^hz_{23}^hz_{31}^h}-\frac{f_{\phi\psi^1\chi^1}}{z_{12}^{h}z_{23}^{h+1}z_{31}^h}-\frac{f_{\psi^1\phi\chi^1}}{z_{12}^{h}z_{23}^{h}z_{31}^{h+1}}\\
\fe
Here, $f_{\phi\phi t^2_t}$ vanishes due to $U(1)_R$ selection rule. By change of variables $t=z_{13}/z_{12}$, \eqref{threeptex} reduces to
\ie\label{twoeq}
(-f_{\phi\phi t_0}-2f_{\psi^1\phi\chi^2}+2f_{\phi\phi\phi}(h_1-h_2+h_3))+t(-2f_{\phi\psi^1\chi^2}+2f_{\psi^1\phi\chi^2}-4f_{\phi\phi\phi}h_3)=0
\fe
As this should be satisfied for all $t>0$, \eqref{twoeq} is equivalent to
\ie
-f_{\phi\phi t_0}-2f_{\psi^1\phi\chi^2}+2f_{\phi\phi\phi}(h_1-h_2+h_3)=0,\quad-2f_{\phi\psi^1\chi^2}+2f_{\psi^1\phi\chi^2}-4f_{\phi\phi\phi}h_3=0
\fe
In this way, by constructing a linear system using all possible non-trivial 3-point function equations, we can solve all $f_{\cF_1\cF_2\cF_3}$ in terms of 10 independent constants. 5 come from $\cF_3\in\cL_0$: $\{f_{\phi\phi\phi},~f_{\phi\phi\tau},~f_{\phi\phi\bar{\tau}},~f_{\phi\phi t_0},~f_{\phi\phi d}\}$, 4 from $\cF_3\in\cL_1$: $\{f_{\phi\phi\psi},~f_{\phi\phi\chi},~f_{\phi\phi C},~f_{\phi\phi\bar{C}}\}$ and 1 from $\cF_3\in\cL_2$: $\{f_{\phi\phi t}\}$. This means that all 3-point OPE coefficients can be expressed in terms of $f_{\phi\phi f}$, where $\phi$ is a superconformal primary, and $f\in\cL_r$. The counting matches with superspace computation in \secref{subsec:4ptinv}. Moreover, the solution set is unique. This provides a strong consistency check of this rather tedious computation. We could check explicitly that all 3-point functions with $\cF_3\in\cL_r$ with $r>2$ vanish, which is not surprising due to the R-symmetry selection rule. We have seen this pattern in the short-multiplet analysis \eqref{short} too. Hence, we can focus on $\cF_3\in\{\cL_0,\cL_1,\cL_2\}$ from now on.
\subsection{4-point functions}\label{subsec:4pt1}
From the above computation, we have gathered all information to construct 4-point function defined in \eqref{scb}. It remains then to find the independent set of external 4-points. The strategy is the same as that of lower correlators. Instead, we stop after solving $T^+_0$ equations, which give a set of independent 4-point functions. We could proceed to solve $G$, $\bar{G}$ equations to produce crossing equations, but we can equivalently construct all the 4-point functions as described at the beginning of this section \secref{sec:SCBlock} only using 2-point and 3-point functions and also crossing equations that we will describe in the next subsection \secref{subsec:cross}.

We construct a linear system of equation commuting $T^+_0$ inside 4-point functions.
\ie
0=\la\cF_1\cF_2\cF_3\cF_4T^+_0\ra&=\la\cF_1\cF_2\cF_3[T^+_0,\cF_4]\ra+\la\cF_1\cF_2[T^+_0,\cF_3]\cF_4\ra+\la\cF_1[T^+_0,\cF_2]\cF_3\cF_4\ra\\
&+\la [T^+_0,\cF_1]\cF_2\cF_3\cF_4\ra
\fe
with 
\ie\label{fptfunction}
\la\cF_1\cF_2\cF_3\cF_4\ra=\frac{f_{1234}[z]}{z_{12}^{h_1+h_2}z_{34}^{h_3+h_4}}\bigg(\frac{z_{24}}{z_{14}}\bigg)^{h_1-h_2}\bigg(\frac{z_{14}}{z_{13}}\bigg)^{h_3-h_4}
\fe
Note that different from before, 4-point function coefficient $f_{1234}[z]$ is not a constant, but a function of cross-ratio $z=\frac{z_{12}z_{34}}{z_{13}z_{24}}$. However, it will not make things complicated, as $T^+_0$ action does not generate any $z_i$ dependence $C(\pa_{z_i},z_i)$. After solving all $T^+_0$ equations, we get $4826$ 4-point functions that will give non-trivial equations from $G$ or $\bar{G}$ commutations.

By using superconformal invariance, we can fix 4-point function of long-multiplet $\la\cL_0\cL_0\cL_0\cL_0\ra$ with first two to be $\cL_0$ but last two operators to be superconformal primary $\phi\in\cL_0$. In other words, in a particular frame: $z_3\ria0$, $z_4\ria\infty$, $\theta_3,\theta_4,\bar{\theta}_3,\bar{\theta}_4\ria0$, $\la\Phi_1\Phi_2\Phi_3\Phi_4\ra$ reduces to $\la\Phi_1\Phi_2\phi_3\phi_4\ra$. If expanded in components, it is a linear combination of $16\times 16=256$ different component 4-point functions. However, they are not all independent, due to the superconformal symmetry. 

With the above $T^+_0$ equations, the $256$ equations reduce into $42$ independent 4-point functions that are
\ie\label{42}
&\text{Total Level 0: }\{{f}_0=\la\phi\phi\phi\phi\ra\},\\
&\text{Total Level 1: }\{{f}_1=\la\psi^1\chi^2\phi\phi\ra,~~{f}_2=\la\chi^1\psi^2\phi\phi\ra,~~{f}_3=\la\phi t_0\phi\phi\ra,~~{f}_4=\la t_0\phi\phi\phi\ra,~~{f}_5=\la \psi^1\psi^2\phi\phi\ra,\\
&{f}_6=\la\phi \bar{\tau}\phi\phi\ra,~~{f}_7=\la \bar{\tau}\phi\phi\phi\ra,~~{f}_8=\la\chi^1\chi^2\phi\phi\ra,~~{f}_9=\la\phi \tau\phi\phi\ra,~~{f}_{10}=\la \tau\phi\phi\phi\ra\},\\
&\text{Total Level 2: }\{{f}_{11}=\la t_{0}t_{0}\phi\phi\ra,~~{f}_{12}=\la t_{0}\tau\phi\phi\ra,~~{f}_{13}=\la t_{0}\bar{\tau}\phi\phi\ra,~~{f}_{14}=\la \tau t_{0}\phi\phi\ra,~~{f}_{15}=\la \tau\tau\phi\phi\ra\\
&{f}_{16}=\la \tau\bar{\tau}\phi\phi\ra,~~{f}_{17}=\la\phi \bar{\tau}t_{0}\phi\ra,~~{f}_{18}=\la \bar{\tau}\tau\phi\phi\ra,~~{f}_{19}=\la\bar{\tau}\bar{\tau}\phi\phi\ra,~~{f}_{20}=\la\psi^1\bar{C}^2\phi\phi\ra,\\
&{f}_{21}=\la\psi^1{C}^2\phi\phi\ra,~~{f}_{22}=\la\chi^1\bar{C}^2\phi\phi\ra,~~{f}_{23}=\la\chi^1{C}^2\phi\phi\ra,~~{f}_{24}=\la\bar{C}^1\psi^2\phi\phi\ra,~~{f}_{25}=\la{C}^1\psi^2\phi\phi\ra\\
&{f}_{26}=\la\bar{C}^1\chi^2\phi\phi\ra,~~{f}_{27}=\la{C}^1\chi^2\phi\phi\ra,~~{f}_{28}=\la d\phi\phi\phi\ra,~~{f}_{29}=\la \phi d\phi\phi\ra,~~{f}_{30}=\la t_{t}^1t_{t}^3\phi\phi\ra\}\\
&\text{Total Level 3: }\{{f}_{31}=\la dt_0\phi\phi\ra,~~{f}_{32}=\la d\tau\phi\phi\ra,~~{f}_{33}=\la D\bar{\tau}\phi\phi\ra,~~{f}_{34}=\la t_{0}d\phi\phi\ra,~~{f}_{35}=\la \tau d\phi\phi\ra,\\
&{f}_{36}=\la \bar{\tau}d\phi\phi\ra,~~{f}_{37}=\la C^1_1C_2\phi\phi\ra,~~{f}_{38}=\la \bar{C}^1C^2\phi\phi\ra,~~{f}_{39}=\la C^1\bar{C}^2\phi\phi\ra,~~{f}_{40}=\la\bar{C}^1\bar{C}^2\phi\phi\ra\},\\
&\text{Total Level 4: }\{f_{41}=\la dd\phi\phi\ra\}
\fe
Here, we classified 4-point functions by the sum of level of 4 external operators. The same set of 42 independent 4-point function will appear in the superspace derivation \eqref{single},\eqref{two},\eqref{three},\eqref{four}. 

As described earlier in the section, with the above 2-point, 3-point function data, one can compute the conformal block expansion of one of 4-point functions $f_i$. Before that, let us rename 10 independent 3-point coefficients $\{f_{\phi\phi\cF}\}$ as
\ie
\cF\in\cL_0,&\quad\{f_{\phi\phi\phi},~f_{\phi\phi\tau},~f_{\phi\phi\bar{\tau}},~f_{\phi\phi t_0},~f_{\phi\phi d}\}=\{a[1],a[2],a[3],a[4],a[5]\}\\
\cF\in\cL_1,&\quad\{f_{\phi\phi\psi},~f_{\phi\phi\chi},~f_{\phi\phi C},~f_{\phi\phi\bar{C}}\}=\{a[6],a[7],a[8],a[9]\}\\
\cF\in\cL_2,&\quad\{f_{\phi\phi t}\}=\{a[10]\}
\fe
and choose 2-point function normalization $f_{\phi\phi}=1$.

Consider for example, $\la\tau\phi\phi\phi\ra$. By \eqref{4ptf}, it is
\ie
&\la\tau\phi\phi\phi\ra=(1-z)^{2h}\bigg[\bigg(\frac{f_{\tau\phi\phi}f_{\phi\phi\phi}}{f_{\phi\phi}}g^{1,0}_0[z]+\frac{f_{\tau\phi\bar{\tau}}f_{\phi\phi\tau}}{f_{\tau\bar{\tau}}}g^{1,0}_1[z]+\frac{f_{\tau\phi\tau}f_{\bar{\tau}\phi\phi}}{f_{\tau\bar{\tau}}}g^{1,0}_1[z]+\frac{f_{\tau\phi t_0}f_{t_0\phi\phi}}{f_{t_0t_0}}g^{1,0}_1[z]\\
&+\frac{f_{\tau\phi d}f_{d\phi\phi}}{f_{dd}}g^{1,0}_2[z]\bigg)_0+\bigg(\frac{f_{\tau\phi\psi}f_{\chi\phi\phi}}{f_{\psi\chi}}g^{1,0}_{\frac{1}{2}}[z]+\frac{f_{\tau\phi\chi}f_{\psi\phi\phi}}{f_{\chi\psi}}\}g^{1,0}_{\frac{1}{2}}[z]+\frac{f_{\tau\phi C}f_{\bar{C}\phi\phi}}{f_{C\bar{C}}}g^{1,0}_{\frac{3}{2}}[z]+\frac{f_{\tau\phi\bar{C}}f_{C\phi\phi}}{f_{\bar{C}C}}g^{1,0}_{\frac{3}{2}}[z]\bigg)_1\\
&+\bigg(\frac{f_{\tau\phi t}f_{t\phi\phi}}{f_{tt}}g^{1,0}_1[z]\bigg)_2\bigg]
\fe
where the subscripts under the big parenthesis denote the rank r of $\cL_r$ to which the exchange primary operator belongs. By using the 3-point, 2-point function solution, it can be expressed in terms of $\{a[1],\ldots,a[10]\}$:
\ie\label{scbex}
&\la\tau\phi\phi\phi\ra=(1-z)^{2h+1}\bigg(a[1]a[2](g^{1,0}_0[z]+\frac{1+h}{2(1+2h)}g^{1,0}_1[z])+a[2]a[5](-\frac{1}{4h(h+1)}g^{1,0}_1[z]\\
&-\frac{h+2}{8h(h+1)(3+2h)}g^{1,0}_2[z])+a[6]^2\frac{1}{4}g^{1,0}_{\frac{1}{2}}[z]+a[6]a[8](\frac{1}{2(3+2h)}g^{1,0}_{\frac{1}{2}}[z]-\frac{1}{16(1+h)}g^{1,0}_{\frac{3}{2}}[z])\\
&+a[8]^2\frac{1}{8(3+5h+2h^2}g^{1,0}_{\frac{3}{2}}[z]\bigg)
\fe
\subsection{Crossing equations}\label{subsec:cross}
By exchanging $\cF_1$ and $\cF_3$ in $\la\cF_1\cF_2\cF_3\cF_4\ra$, we get crossing channel $\la\cF_3\cF_2\cF_1\cF_4\ra$. As we know all possible $\la\cF_3\cF_2\cO\ra$, $\la\cO'\cF_1\cF_4\ra$, $\la\cO\cO'\ra$, we can compute all 42 crossing channel superconformal blocks that correspond to \eqref{42}.
\ie
\la \cF_1\cF_3\cF_2\cF_4\ra&=\frac{1}{x^{2\Delta_{\cF_1}}_{13}x^{2\Delta_{\cF_2}}_{24}}\bigg(\sum_{\{\cO_1\}} \frac{f_{\cF_1\cF_3\cO_1}f_{\cF_2\cF_4\cO'_1}}{f_{\cO_1\cO'_1}}g_{h_{\cO_1}}(1-z)+\ldots+\sum_{\{\cO_i\}}\frac{f_{\cF_1\cF_3\cO_n}f_{\cF_2\cF_4\cO'_n}}{f_{\cO_n\cO'_n}}g_{h_{\cO_n}}(1-z)\bigg)
\fe
For instance, $1\leftrightarrow3$ crossing channel of \eqref{scbex} is
\ie\label{crossex}
&\la\phi\phi\tau\phi\ra=z^{1+2h}\bigg(a[1]a[2](g^{1,0}_0[1-z]+\frac{1+h}{2+4h}g^{1,0}_1[1-z])-a[2]a[5](\frac{6+4h}{8h(1+h)(3+2h)}g^{1,0}_1[1-z]\\
&+\frac{2+h}{8h(1+h)(3+2h)}g^{1,0}_2[1-z])-a[6]^2(\frac{1}{4}g^{1,0}_{\frac{1}{2}}[1-z])+a[6]a[8](\frac{1}{6+4h}g^{1,0}_{\frac{1}{2}}[1-z]\\
&+\frac{1}{16(1+h)}g^{1,0}_{\frac{3}{2}}[1-z])+a[8]^2(\frac{1}{8(3+5h+2h^2)}g^{1,0}_{\frac{3}{2}}[1-z])\bigg)
\fe
We have dropped anti-holomorphic part of equations until now, and now we want to restore it. Since we assume that only right moving part is $\cN=4$ supersymmetric, we can simply replace $z$ dependent factors or functions with following rules:
\ie
z^{i+2h}\ria z^{i+2h}\bar{z}^{2\bar{h}},\quad g^{\Delta_{12},\Delta_{34}}_{h_{ex}}[z]\ria g^{\Delta_{12},\Delta_{34}}_{h_{ex}}[z,\bar{z}],\quad g^{\Delta_{12},\Delta_{34}}_{h_{ex}}[1-z]\ria g^{\Delta_{12},\Delta_{34}}_{h_{ex}}[1-z,1-\bar{z}]
\fe 
where $\Delta=\frac{h+\bar{h}}{2}$.

By equating $\la\cF_1\cF_2\cF_3\cF_4\ra_n$ and $\la\cF_3\cF_2\cF_1\cF_4\ra_n$ for each $n=1,\ldots,42$, we arrive at a system of 42 linear equations that can be represented by fourty two $10\times10$ block diagonal $F^n_{ij}$ matrices.
\ie\label{cross}
\sum_{i,j=1}^{10} a[i]F^n_{ij}(z)a[j]=0,\quad n=1,\ldots,42
\fe
Most of the matrix component of $F^n_{ij}$ are zero, as one can see in \eqref{scbex}, \eqref{crossex}. There are 195 independent crossing equations that we need to solve using SDPB. We provide selected few in the Appendix and the complete set of crossing equations is available in the separate Mathematica file.

\section{$\cN=4$ superspace approach}\label{sec:superspace}
In this section, we explain a separate approach to analyze $\cN=4$ long-multiplet 4-point functions using the superspace and the Casimir differential equations, generalizing the $\cN=2$ superspace approach that was introduced in \cite{Cornagliotto:2017dup}. We have obtained 3-point, 4-point superconformal invariants, and Nilpotent invariants that are used in the long-multiplet 4-point function expansion and the Casimir differential operator that can be used to get the conformal block expansion. Due to a subtle problem in $\cN=4$ superspace, we could not get the final expression for superconformal blocks, but we proceeded as much as possible and pointed out the problem. 

In this section, we heavily used Mathematica package `grassmann.m' developed by Matthew Headrick \cite{Headrick}. For concise presentation, we will drop left-moving non-supersymmetric part of 4-point functions consistently throughout the section and re-introduce in the appropriate place.

We want to study long-multiplet $\cL_0$ 4-point function $\langle\Phi_1(Z_1)\Phi_2(Z_2)\Phi_3(Z_3)\Phi_4(Z_4)\rangle$, with $Z_i=(z_i,\theta_i,\bar{\theta}_i)$. In $\cN=4$ superspace, a generic long multiplet is represented as
\ie\label{super4}
\Phi(x,\theta^\A,\bar{\theta}^\A)=\phi(x)+\theta\psi(x)+\bar{\theta}\chi(x)+\theta\bar{\theta}t_0(x)+\theta\theta\bar{\tau}+\bar{\theta}\bar{\theta}\tau+\theta\sigma^i\bar{\theta}t^i_t+\theta\bar{\theta}\theta\bar{C}+\theta\bar{\theta}C\bar{\theta}+(\theta\bar{\theta})^2d
\fe
with $SU(2)_R$ index all contracted. Quantum numbers for each element are $(h,0)_\phi$, $(h+1/2,1/2)_{\psi^\A,\chi^\A}$, $(h+1,0)_{\tau,\bar{\tau},t_0}$, $(h+1,1)_{t^i_t}$,$(h+3/2,1/2)_{C^\A,\bar{C}^\A}$, $(h+2,0)_D$. With the explicit superspace expansion \eqref{super4}, one can expand the 4-point function in terms of nilpotent superconformal invariants $\{\cI_i,\cJ_j,\cK_k\}$ that we will derive in this section, as
\ie\label{cb}
\la\Phi_1(Z_1)\Phi_2(Z_2)\Phi_3(Z_3)\Phi_4(Z_4)\ra=\sum_{i,j,k}g_n(\cI_i,\cJ_j,\cK_k)F_n(z)
\fe
where $g_n$ is a monomial of $\{\cI_i,\cJ_j,\cK_k\}$ and $F_n(z)$ is component 4-point function such as $\la\psi^1\chi^2\phi\phi\ra$. By studying $\cN=4$ superspace 3-point invariants $U_{123}$ and 4-point invariants $\{\cI_i,\cJ_j,\cK_k\}$, one can systematically deduce the expansion. 

Each of 4-point function $F_n(z)$ can be decomposed into Virasoro conformal blocks labeled by the exchanged conformal primary in one of three long-multiplets: $\cL_0$, $\cL_1$, $\cL_2$.
\ie\label{4.3}
F_n(z)=\sum_{i=1}^{5}c^i_ng^{h_{12},h_{34}}_{h^i_{ex}}(z)
\fe
Here, $h_{12}=h_1-h_2$, $h_{34}=h_3-h_4$, and $h_{ex}$ is weight of exchange primary. Note that we sum 5 terms as there are 5 different levels in a given long-multiplet $\cL_r$-- see the diamond graphs \ref{diamondgeneral}. $g^{h_{12},h_{34}}_{h^i_{ex}}(z)$ is the Virasoro block, derived recursively from $sl(2)$ block.

So, in the superspace approach, there are two things to compute to get crossing equations eventually: 1. the superspace expansion of long-multiplet 4-point function in terms of the superconformal invariants. 2. expansions of each of 4-point functions into Virasoro blocks; in other words we need to get the coefficients $c^i_n$.
\subsection{$\cN=4$ superspace and 3-point invariants}\label{subsec:3ptinv}
2d $\cN=4$ superspace has 4 pairs of grassmann coordinates $\theta_{1,2,3,4}$, $\bar{\theta}_{1,2,3,4}$ along with the usual spacetime coordinates $(z,\bar{z})$. The symmetry that rotates $\theta_i$'s is then $O(4)$. Restricting R-symmetry as $su(2)_R$ subalgebra of $o(4)$ leads to $SU(2)-$extended $\cN=4$ superspace where small $\cN=4$ superconformal algebra is properly embedded. Coordinate of the superspace is then $Z=(z,\theta^\A,\bar{\theta}^\A)$, where $\theta^\A$ and $\bar{\theta}^\A$ are $2$ and $\bar{2}$ under $SU(2)_R$. One can then introduce (super-)translation invariants that are building blocks for $n-$point invariants.
\ie
Z_{ij}=z_i-z_j-\theta_j\bar{\theta}_i+\theta_i\bar{\theta}_j,\quad\theta_{ij}=\theta_i-\theta_j,\quad\bar{\theta}_{ij}=\bar{\theta}_i-\bar{\theta}_j
\fe

We derived 3-point superspace invariants that are function of three superspace coordinates, and are invariant under all superconformal transformations. If one starts from ansatz that only depends on (super)translation invariants, the main task is to impose an inversion invariance that guarantees conformal invariance. We present the detail of the derivation in Appendix \secref{app:integrand}. The 3-point invariants are
\ie\label{3ptinvariants}
U_{123}&=\frac{Z_{23}^2\theta_{13}\bar{\theta}_{13}-(\theta_{13}\bar{\theta}_{13}+\theta_{23}\bar{\theta}_{23}-\theta_{12}\bar{\theta}_{12})Z_{13}Z_{23}+\theta_{23}\bar{\theta}_{23}Z_{13}^2}{Z_{13}Z_{23}Z_{12}}\\
V_{123}&=\frac{Z_{23}^2\theta_{13}{\theta}_{13}-(\theta_{13}{\theta}_{13}+\theta_{23}{\theta}_{23}-\theta_{12}{\theta}_{12})Z_{13}Z_{23}+\theta_{23}{\theta}_{23}Z_{13}^2}{Z_{13}Z_{23}Z_{12}}\\
W_{123}&=\frac{Z_{23}^2\bar{\theta}_{13}\bar{\theta}_{13}-(\bar{\theta}_{13}\bar{\theta}_{13}+\bar{\theta}_{23}\bar{\theta}_{23}-\bar{\theta}_{12}\bar{\theta}_{12})Z_{13}Z_{23}+\bar{\theta}_{23}\bar{\theta}_{23}Z_{13}^2}{Z_{13}Z_{23}Z_{12}}\\
\fe
\subsection{4-point invariants and their limits}\label{subsec:4ptinv}
Nine 4-point invariants that consist of fermionic bilinears are obtained by replacing indices $\{123\}$ of \eqref{3ptinvariants}, to $\{124\},\{134\},\{234\}$. With the usual bosonic 4-point invariant $U_1$ and its fermionic partner $U_5$, we complete eleven 4-point invariants. From now, we will use following definitions.
\ie
U_1:=\frac{Z_{13}Z_{24}}{Z_{23}Z_{14}},\quad &U_2:=U_{124},\quad U_3:=U_{134},\quad U_4:=U_{234},\quad U_5=\frac{Z_{12}Z_{34}}{Z_{23}Z_{14}},\\
&V_2:=V_{124},\quad V_3:=V_{134},\quad V_4:=V_{234},\\
&W_2:=W_{124},\quad W_3:=W_{134},\quad W_4:=W_{234}
\fe
Hence, there are 10 four-point invariants constructed from fermionic bilinears and 1 four-point invariant from usual bosonic coordinates in $\cN=4$ superspace. The number 10 matches with the number of independent 3-point OPE coefficients obtained in the previous section \secref{subsec:3pt}.

Due to grassmann nature, $\{U_i,V_j,W_k\}$ are nilpotent. This is the reason that one can use those invariants when expanding long-multiplet 4-point functions as it guarantees finite truncation in the superspace expansion. To obtain clear nilpotency relations, we want to convert the basis into a special form. To guess the form of the nilpotent invariants, first let us take following limits of the 4-point invariants: $x_4\ria\infty$, $x_3\ria0$, $\theta_3,\theta_4\ria0$, $\bar{\theta}_3,\bar{\theta}_4\ria0$.
\ie\label{4pt}
U_1&\ria\frac{z_1}{z_2},\quad U_2\ria\frac{\theta_1\bar{\theta}_1-\theta_1\bar{\theta}_2-\theta_2\bar{\theta}_1+\theta_2\bar{\theta}_2}{z_1-z_2-\theta_1\bar{\theta}_2+\theta_2\bar{\theta}_1},\quad U_3\ria\frac{\theta_1\bar{\theta}_1}{z_1},\quad U_4\ria\frac{\theta_2\bar{\theta}_2}{z_2},\quad U_5\ria\frac{z_1-z_2-\theta_1\bar{\theta}_2+\theta_2\bar{\theta}_1}{z_2}\\
V_2&\ria\frac{\theta_1{\theta}_1-\theta_1{\theta}_2-\theta_2{\theta}_1+\theta_2{\theta}_2}{z_1-z_2-\theta_1\bar{\theta}_2+\theta_2\bar{\theta}_1},\quad V_3\ria\frac{\theta_1{\theta}_1}{z_1},\quad V_4\ria\frac{\theta_2{\theta}_2}{z_2}\\
W_2&\ria\frac{\bar{\theta}_1\bar{\theta}_1-\bar{\theta}_1\bar{\theta}_2-\bar{\theta}_2\bar{\theta}_1+\bar{\theta}_2\bar{\theta}_2}{z_1-z_2-{\theta}_1\bar{\theta}_2+{\theta}_2\bar{\theta}_1},\quad W_3\ria\frac{\bar{\theta}_1\bar{\theta}_1}{z_1},\quad W_4\ria\frac{\bar{\theta}_2\bar{\theta}_2}{z_2}
\fe
\subsection{Nilpotent invariants and their independent combinations}\label{subsec:Nilpotentinv}
From \eqref{4pt}, we get some hint to construct the good basis of the nilpotent invariants $\{\cI_i,\cJ_i,\cK_i\}$. The invariants defined in \eqref{4pt} should combine to produce simple limits. Hence, we can guess following combinations and compute their limit in the convenient frame.
\ie\label{nilpotent}
I_0&=U_1&&\ria~\cI_0:=\frac{z_1}{z_2}\\
I_1&=-U_5+U_1-1&&\ria~\cI_1:=\frac{\theta_1\bar{\theta}_2-\theta_2\bar{\theta}_1}{z_2}\\
I_2&=-U_2U_5+U_3U_1+U_4&&\ria~\cI_2:=\frac{\theta_1\bar{\theta}_2+\theta_2\bar{\theta}_1}{z_2}\\
I_3&=U_4&&\ria~\cI_3:=\frac{\theta_2\bar{\theta}_2}{z_2}\\
I_4&=U_3U_1&&\ria~\cI_4:=\frac{\theta_1\bar{\theta}_1}{z_2}\\
J_2&=-V_2U_5+V_3U_1+V_4&&\ria~\cJ_2:=\frac{\theta_1\theta_2+\theta_2\theta_1}{z_2}=\frac{2\theta_1\theta_2}{z_2}\\
J_3&=V_4&&\ria~\cJ_3:=\frac{\theta_2\theta_2}{z_2}\\
J_4&=V_3U_1&&\ria~\cJ_4:=\frac{\theta_1\theta_1}{z_2}\\
K_2&=-W_2U_5+W_3U_1+W_4&&\ria~\cK_2:=\frac{\bar{\theta}_1\bar{\theta}_2+\bar{\theta}_2\bar{\theta}_1}{z_2}=\frac{2\bar{\theta}_1\bar{\theta}_2}{z_2}\\
K_3&=W_4&&\ria~\cK_3:=\frac{\bar{\theta}_2\bar{\theta}_2}{z_2}\\
K_4&=W_3U_1&&\ria~\cK_4:=\frac{\bar{\theta}_1\bar{\theta}_1}{z_2}
\fe
As the nilpotency condition preserves under the conformal transformations, we can use $\{\cI_i,\cJ_i,\cK_i\}$ to figure out the whole expansion of the long-multiplet 4-point function. In this frame, \eqref{cb} reduces to
\ie\label{spec}
\la\Phi_1\Phi_2\phi_3\phi_4\ra=\sum_{i,j,k}g_n(\cI_i,\cJ_j,\cK_k)F_n(z)
\fe

Now, we need to get all independent $g_n(\cI_i,\cJ_j,\cK_k)$. We start by writing down all possible letters and words and reduce the set by using algebraic relations between them. Some obvious nilpotent relations are following. From now, we will redefine $\cI_1=(\cI_1+\cI_2)/2$ and $\cI_2=(\cI_1-\cI_2)/2$.
\ie
&\cI_1^3=\cI_2^3=\cI_3^3=\cI_4^3=\cJ_2^3=\cJ_3^2=\cJ_4^2=\cK_2^3=\cK_3^2=\cK_4^2=0\\
&\cI_1\cI_4^2=\cI_1\cI_3^2=\cI_2\cI_3^2=\cI_2\cI_4^2=0\\
&\cI_3\cI_1^2=\cI_3\cI_2^2=\cI_4\cI_1^2=\cI_4\cI_2^2=0\\
&\cI_1\cJ_4=\cI_1\cK_3=\cI_2\cJ_3=\cI_2\cK_4=\cI_3\cJ_3=\cI_3\cK_3=\cI_4\cJ_4=\cI_4\cK_4=0\\
&\cK_2\cK_3=\cK_2\cK_4=\cJ_2\cJ_3=\cJ_2\cJ_4=0
\fe
It is also possible to deduce all non-vanishing $g(\cI_i,\cJ_j,\cK_k)$. We will classify them by the number of letters. 
\ie\label{total}\nonumber
&\text{Single Letter : }\cI_1,~\cI_2,~\cI_3,~\cI_4,~\cJ_2,~\cJ_3,~\cJ_4,~\cK_2,~\cK_3,~\cK_4\\
&\text{Two Letters : }\cI_1^2,~\cI_1\cI_2,~\cI_1\cI_3,~\cI_1\cJ_2,~\cI_1\cJ_3,~\cI_1\cK_2,~\cI_1\cK_4,~\cI_2^2,~\cI_2\cI_3,~\cI_2\cI_4,~\cI_2\cJ_2,\\
&\cI_2\cJ_4,~\cI_2\cK_2,~\cI_2\cK_3,~\cI_3\cI_4,~\cI_3\cJ_4,~\cI_3\cJ_2,~\cI_3\cK_4,~\cI_3\cK_2,~\cI_4^2,~\cI_4\cK_3,~\cI_4\cK_2,~\cI_4\cJ_3,~\cI_4\cJ_2\\
&\cJ_2^2,~\cJ_2K_2,~\cJ_2K_3,~\cJ_2K_4,~\cJ_3K_4,~\cJ_3K_3,~\cJ_3\cJ_4,~\cJ_4K_2,~\cJ_4K_4,~K_2^2,~K_3K_4\\
&\text{Three Letters : }\cI_1^2\cI_2,~\cI_1\cI_3\cI_4,~\cI_1\cJ_2 K_2,~\cI_2 \cJ_4 K_3,~\cI_3 \cJ_4 K_2,~\cI_4 \cJ_2 K_3,~\cI_1^2 \cJ_3,~\cI_1 \cI_3 \cJ_2,~\cI_3^2 \cJ_4\\
&\cJ_2^2 K_3,~\cJ_3 \cJ_4 K_3,~\cI_1^2 K_4,~\cI_1 \cI_4 K_2,~\cI_4^2 K_3,~\cJ_4 K_2^2,~\cJ_4 K_3 K_4,~\cI_1 \cI_2^2,~\cI_1 \cJ_3 K_4,~\cI_2 \cI_3 \cI_4,~\cI_2 \cJ_2 K_2\\
&\cI_3 \cJ_2 K_4,~\cI_4 \cJ_3 K_2,~ \cI_1 \cI_2 \cI_3,~\cI_1 \cJ_3 K_2,~\cI_2 \cJ_2 K_3,~\cI_3^2 \cI_4,~\cI_3 \cJ_2 K_2,~\cI_4 \cJ_3 K_3,~\cI_1 \cI_2 \cI_4,~\cI_1 \cJ_2 K_4,~\cI_2 \cJ_4 K_2\\
&\cI_3 \cI_4^2,~\cI_3 \cJ_4 K_4,~\cI_4 \cJ_2 K_2,~ \cI_1 \cI_2 \cJ_2,~\cI_1 \cI_4 \cJ_3,~\cI_2 \cI_3 \cJ_4,~\cI_3 \cI_4 \cJ_2,~\cJ_2^2 K_2,~\cJ_3 \cJ_4 K_2,~ \cI_1 \cI_2 K_2,~\cI_1 \cI_3 K_4\\
&\cI_2 \cI_4 K_3,~\cI_3 \cI_4 K_2,~\cJ_2 K_2^2,~\cJ_2 K_3 K_4,~ \cI_2^2 \cJ_4,~\cI_2 \cI_4 \cJ_2,~\cI_4^2 \cJ_3,~\cJ_2^2 K_4,~\cJ_3 \cJ_4 K_4,~ \cI_2^2 K_3,~\cI_2 \cI_3 K_2\\
&\cI_3^2 K_4,~\cJ_3 K_2^2,~\cJ_3 K_3 K_4\\
&\text{Four Letters: }\cI_1^2\cI_2^2,~\cI_3^2\cI_4^2,~\cJ_2^2K_2^2,~\cJ_3\cJ_4K_3K_4
\fe
\quad First of all, single-letters are all independent; we can not express any of those in terms of a linear combination of the others. There are {\bf{10}} of them.
\ie\label{single}
\{\cI_1,~\cI_2,~\cI_3,~\cI_4,~\cJ_2,~\cJ_3,~\cJ_4,~\cK_2,~\cK_3,~\cK_4\}
\fe
There are many two-letters relations between the invariants, part of which we wrote down below:
\ie\label{secondrelations}
&\{\cI_1\cI_2+\cI_3\cI_4+\cJ_2K_2=0,~~
2\cI_1\cI_3+\cJ_2K_3=0,~~
2\cI_1\cJ_2+\cI_3\cJ_4=0,~~
\cI_1\cJ_3+2\cI_3\cJ_2=0,\\
&2\cI_1K_2+\cI_4K_3=0,~~\cI_1K_4+2\cI_4K_2=0,~~
2\cI_2\cI_3+\cJ_3K_2=0,~~
2\cI_2\cI_4+\cJ_2K_4=0,~~
2\cI_2\cJ_2+\cI_4\cJ_3=0\}
\fe
These relations reduce the number of two letters from 39 to {\bf{20}}. 
\ie\label{two}
&\{(\cI_1)^2,~~\cI_1 \cI_2,~~\cI_3 \cI_4,~~\cI_1 \cI_3,~~\cI_1 \cI_4,~~\cI_1 \cJ_2,~~\cI_1 \cJ_3,~~\cI_1 K_2,~~\cI_1 K_4,~~(\cI_2)^2,~~\cI_2 \cI_3\\
&\cI_2 \cI_4,~~\cI_2 \cJ_2,~~\cI_2 \cJ_4,~~\cI_2 K_2,~~\cI_2 K_3,~~\cI_3^2,~~\cI_4^2,~~\cJ_2^2,~~K_2^2\}
\fe
Using three-letters relations
\ie
(\cI_1)^2\cI_2+2\cI_1\cI_3\cI_4=\cI_1(\cI_2)^2+2\cI_2\cI_3\cI_4=2\cI_1\cI_2\cI_3+\cI_3^2\cI_4=2(\cI_2)^2\cI_4+\cI_3\cI_4^2=0
\fe
we can also reduce the number of three letters from 56 to {\bf{10}} that are
\ie\label{three}
&(\cI_1)^2\cI_2,~~(\cI_1)^2\cJ_3,~~(\cI_1)^2 K_4,~~\cI_1(\cI_2)^2,~~\cI_1\cI_2\cI_3,~~\cI_1\cI_2\cI_4,~~\cI_1\cI_2\cJ_2,~~\cI_1 \cI_2 K_2,~~(\cI_2)^2 \cJ_4,~~(\cI_2)^2 K_3
\fe
Trivially, there is {\bf{1}} independent four-letter:
\ie\label{four}
\cI_1^2\cI_2^2
\fe
Hence, including the bosonic single letter $\cI_0$, the total number of the independent combinations of the nilpotent invariants is $1+10+20+10+1=42$, which matches the counting from the previous section \secref{sec:SCBlock}, \eqref{42}. It must be the linearly independent set, since each of 42 combinations has different number of $(\theta_1,\theta_2,\bar{\theta}_1,\bar{\theta}_2)$. We further checked those of $\cV$ are all independent. Let us call the set of 42 combinations of invariants $\cS$ and their general element $\cS_i$
\subsection{Crossing Equations}\label{subsec:crossingeq}
To write down the crossing equations, we first need to derive the crossing transformed invariants. The crossing acts on $\{\cI_i,\cJ_j,K_j\}$ by exchanging $(z_1,\bar{z}_1,\theta_1,\bar{\theta}_1)$ and $(z_3,\bar{z}_3,\theta_3,\bar{\theta}_3)$. The crossing symmetry imposes following constraint:
\ie\label{spec2}
\sum_{i}\cS_i F_i(z)\propto\sum_{i}{\cS^t_i}F_i(1-z)
\fe
The RHS of \eqref{spec2} can be rearranged into an expansion with the same set of parameters of LHS, since we have seen 41 combinations of the nilpotent invariants are linearly independent and span the set of possible 4-point invariants. We could find $\cI^t_\cI$, $\cJ^t_j$, $\cK^t_k$.
\ie\label{crossingtransform}
\cI^t_0&=-\frac{\cI_0}{\cI_1+\cI_2 + 1 - \cI_0},~~\cI^t_1=-\frac{\cI_1+\cI_2}{\cI_1+\cI_2 + 1 - \cI_0},~~\cI^t_2=\frac{-\cI_1+\cI_2+2 \cI_4}{\cI_1+\cI_2 + 1 - \cI_0},~~\cI^t_4=\frac{\cI_4}{\cI_1+\cI_2 + 1 - \cI_0}\\
\cI^t_3&=\frac{\cI_4 + \cI_2 + \cI_3-\cI_1}{\cI_1+\cI_2 + 1 - \cI_0},~~\cJ^t_2=\frac{2 \cJ_4 - \cJ_2}{\cI_1+\cI_2 + 1 - \cI_0},~~ \cJ^t_4=\frac{\cJ_4}{\cI_1+\cI_2 + 1 - \cI_0},~~\cJ^t_3=\frac{\cJ_4 + \cJ_3 - \cJ_2}{\cI_1+\cI_2 + 1 - \cI_0} \\
\cK^t_2&=\frac{2 \cK_4 - \cK_2}{\cI_1+\cI_2 + 1 - \cI_0},~~\cK^t_4=\frac{\cK_4}{\cI_1+\cI_2 + 1 - \cI_0},~~\cK^t_3=\frac{\cK_3 + \cK_4 - \cK_2}{\cI_1+\cI_2 + 1 - \cI_0}
\fe
From this, one can deduce the crossing transformed set of the nilpotent invariants $\{\cS^t_i\}$.

Given the above information, we are ready to write down the crossing equations, starting from $1-2$, $3-4$ channel 4-point function:
\ie
\la\Phi(\cZ_1,\bar{z}_1)\Phi(\cZ_2,\bar{z}_1)\Phi(\cZ_3,\bar{z}_3)\Phi(\cZ_4,\bar{z}_4)\ra=\frac{1}{Z_{12}^{2h}}\frac{1}{Z^{2h}_{34}}\frac{1}{\bar{z}^{2\bar{h}}_{12}\bar{z}^{2\bar{h}}_{34}}\big(g_0(\cI_0,\bar{z})+\sum^{41}_{i=1}\cS_ig_i(\cI_0,\bar{z})\big)
\fe
where $\cS_i\in\cS$. Here we coupled with a left-moving non-supersymmetric conformal block that adds $\bar{z}$ dependence. The crossing channel is 
\ie
\la\Phi(\cZ_3,\bar{z}_3)\Phi(\cZ_2,\bar{z}_2)\Phi(\cZ_1,\bar{z}_1)\Phi(\cZ_4,\bar{z}_4)\ra=\frac{1}{Z_{32}^{2h}}\frac{1}{Z^{2h}_{14}}\frac{1}{\bar{z}^{2\bar{h}}_{32}\bar{z}^{2\bar{h}}_{14}}\big(g_0(\cS^t_0,\bar{z})+\sum^{41}_{i=1}\cS^t_ig_i(\cS^t_0,\bar{z})\big)
\fe
The crossing equation is then
\ie
g_0(\cI_0,\bar{z})+\sum^{41}_{i=1}\cS_ig_i(\cI_0,\bar{z})=(\cI_0-\cI_1-\cI_2-1)^{2h}\bigg(\frac{\bar{z}}{\bar{z}-1}\bigg)^{2\bar{h}}\bigg(g_0(\cS^t_0,1-\bar{z})+\sum^{41}_{i=1}\cS^t_ig_i(\cS^t_0,1-\bar{z})\bigg)
\fe
\subsection{Casimir equation}
Now, it remains to solve $g_n(z,\bar{z})$ that take following form. 
\ie
g_n(z)=c_n^1g^{h_{12},h_{34}}_{h}(z)+c_n^2g^{h_{12},h_{34}}_{h+\frac{1}{2}}(z)+c_n^3g^{h_{12},h_{34}}_{h+1}(z)+c_n^4g^{h_{12},h_{34}}_{h+\frac{3}{2}}(z)+c_n^5g^{h_{12},h_{34}}_{h+2}(z)
\fe
The reason for this particular decomposition is explained around \eqref{4.3}. By solving $g_n(z,\bar{z})$, we mean that we solve for $c^n_i$ with $n=1,\ldots,42$, $i=1,\ldots,5$ using following set of coupled differential equations \cite{Fitzpatrick:2014oza}, which are called Casimir differential equations:
\ie\label{goal}
\cC^{(2)}\begin{pmatrix}g_0\\ g_1 \\ \ldots \\ g_{40}\end{pmatrix} =\cD[\cI_0]\begin{pmatrix}g_0\\ g_1 \\ \ldots \\ g_{40}\end{pmatrix}=c_2\begin{pmatrix}g_0\\ g_1 \\ \ldots \\ g_{40} \end{pmatrix}
\fe
where $\cD[\cI_0]$ is a matrix of differential operators with respect to $\cI_0$ and ${c}_2$ is a $42\times42$ matrix with constant that depends on $h$.

We derived the quadratic Casimir for $\cN=4$. 
\ie\label{scasimir}
\mathcal{C}_2=&\big(L^2_0-\frac{1}{2}\{L_1,L_{-1}\}\big)-\big((T^3_0)^2+\frac{1}{2}\{T^+_0,T^-_0\}\big)+\frac{1}{4}\E_{\A\B}\big(-G^\A_{-\frac{1}{2}}\bar{G}^\B_{\frac{1}{2}}-\bar{G}^\A_{-\frac{1}{2}}{G}^\B_{\frac{1}{2}}+G^\A_{\frac{1}{2}}\bar{G}^\B_{-\frac{1}{2}}+\bar{G}^\A_{\frac{1}{2}}{G}^\B_{-\frac{1}{2}}\big)
\fe
The way to derive it is to start from the most general ansatz $\cC_2=\sum_{i\in{b\cup f}}c_i\cG_i$ that is a linear combination of all possible quadratic global generators that are invariant under the global $\cN=4$ superconformal algebra and fix the coefficients using the algebra, where
\ie
&\text{Quadratic Bosonic Generators: }b=\{L_{\mp}L_{\pm},L_0L_0,T^{\mp}T^{\pm},T^0T^0,L^0T^0\}\\
&\text{Quadratic Fermionic Generators: }f=\{G^i_{-\frac{1}{2}}G^j_{\frac{1}{2}},\bar{G}^i_{-\frac{1}{2}}\bar{G}^j_{\frac{1}{2}},G^i_{\frac{1}{2}}\bar{G}^j_{-\frac{1}{2}},\bar{G}^i_{\frac{1}{2}}{G}^j_{-\frac{1}{2}}\},~i,j=1,2
\fe

After moving to the convenient frame $x_3\ria0$, $x_4\ria\infty$, $\theta_3,\theta_4$, $\bar{\theta}_3,\bar{\theta}_4\ria0$,  the Casimir operators only act on first two operators of 4-point function $\la\Phi_1\Phi_2\phi_3\phi_4\ra$. Hence, we need to get the two particle Casimir operator, similar to \cite{Murugan:2017eto}.
\ie\label{scasimir2}
&\mathcal{C}^{(2)}_{12}=\big(L^{(1)}_0+L^{(2)}_0\big)^2-\frac{1}{2}\{(L^{(1)}_{-1}+L^{(2)}_{-1}),(L^{(1)}_{+1}+L^{(2)}_{+1})\}-\frac{1}{4}\big(\big(T^{(1)}_0+T^{(2)}_0\big)^2-\frac{1}{2}\{(T^{(1)}_{-1}+T^{(2)}_{-1}),(T^{(1)}_{+1}+T^{(2)}_{+1})\}\big)\\
&+\frac{1}{2}\big[\big(\bar{G}^{(1)}_{\frac{1}{2}}+\bar{G}^{(2)}_{\frac{1}{2}}\big),\big(G^{(1)}_{-\frac{1}{2}}+G^{(2)}_{-\frac{1}{2}}\big)\big]+\frac{1}{2}\big[\big(G^{(1)}_{\frac{1}{2}}+G^{(2)}_{\frac{1}{2}}\big),\big(\bar{G}^{(1)}_{-\frac{1}{2}}+\bar{G}^{(2)}_{-\frac{1}{2}}\big)\big]\\
&=2c_2+2L^{(1)}_0L^{(2)}_0-L^{(1)}_{-1}L^{(2)}_{+1}-L^{(1)}_{+1}L^{(2)}_{-1}-\frac{1}{4}\big(2T^{(1)}_0T^{(2)}_0-T^{(1)}_{-1}T^{(2)}_{+1}-T^{(1)}_{+1}T^{(2)}_{-1}\big)\\
&+\bar{G}^{(1)}_{\frac{1}{2}}G^{(2)}_{-\frac{1}{2}}-G^{(1)}_{-\frac{1}{2}}\bar{G}^{(2)}_{\frac{1}{2}}+G^{(1)}_{\frac{1}{2}}\bar{G}^{(2)}_{-\frac{1}{2}}-\bar{G}^{(1)}_{-\frac{1}{2}}G^{(2)}_{\frac{1}{2}}
\fe
Here, the superscripts $(1),(2)$ in the parenthesis refer to first two long-multiplets $\Phi_1$, $\Phi_2$.
\subsection{The puzzle}\label{subsec:puzzle}
To solve the Casimir equation, we need to know the superspace representation of $\cN=4$ superconformal algebra generators that consist of the quadratic Casimir operator \eqref{scasimir}. For simple notation, let us re-introduce small $\cN=4$ superconformal algebra with the outer-automorphism manifest.
The global ${\cal N}=4$ superconformal algebra is
\ie\label{eqn:N=4Commu}
&{[}L_m,L_n]=(m-n)L_{m+n},\quad{[}T_0^i,T_0^j]=i\epsilon^{ijk}T_0^{k},\\
&[L_m,G^{\A A}_{r}]=\left({m\over 2}-r\right)G^{\A A}_{m+r},\quad [T_0^i,G^{\A A}_r]=-{1\over 2}(\sigma^i)_\B{}^\A G^{\B A}_r.
\\
&\{G^{\A A}_{-{1\over 2}},G^{\B B}_{-{1\over 2}}\}=2\epsilon^{\A\B}\epsilon^{AB}L_{-1},
\\
&\{G^{\A A}_{-{1\over 2}},G^{\B B}_{{1\over 2}}\}=2\epsilon^{\A\B}\epsilon^{AB}L_0+2\epsilon^{AB}(\sigma^a)^{\A\B}T_0^a,
\\
&\{G^{\A A}_{{1\over 2}},G^{\B B}_{{1\over 2}}\}=2\epsilon^{\A\B}\epsilon^{AB}L_{-1},
\fe
for $i=1,2,3$, $m,n=0,\pm 1$ and $r=\pm {1\over 2}$. Here, $\A,\B$ indices are that of $SU(2)_F$ outer-automorphism of small $\cN=4$ superconformal algebra.

To find the superspace representation of each generator, we start with the most general ansatz and fix the coefficients $\{p,q,r,s,t,u,v,w,y\}$ in front of each term.
\ie\label{puz}
&L_{-1}=\partial_z
,\quad L_0=z\partial_z+p\theta\partial_\theta,\quad L_{1}=z^2\partial_z +qz\theta\partial_\theta,
\\
&T_0^a=r\theta^{\gamma C}(\sigma^a)_\gamma{}^\D\partial_{\theta^{\D C}},
\\
&G^{\A A}_{-{1\over 2}}=s\epsilon^{\A\B}\epsilon^{AB}\partial_{\theta^{\B B}}+t\theta^{\A A} \partial_z,
\\
&G^{\A A}_{{1\over 2}}=u\epsilon^{\A\B}\epsilon^{AB}z\partial_{\theta^{\B B}}+v\theta^{\A A} \theta^{\B B} \partial_{\theta^{\B B}}+w\theta^{\B A} \theta^{\A B} \partial_{\theta^{\B B}}+y\theta^{\A A}  z\partial_z .
\fe
By using \eqref{eqn:N=4Commu}, we can try to fix the coefficients. However, there is no non-trivial set of solution for the coefficients.\footnote{We thank Carlo Meneghelli for explaining that this problem can be resolved by using more general algebra than \eqref{puz}.} As we did not have a superspace representation of each generator, we could not set up the Casimir differential equation that would solve to coefficients in the conformal block expansions.

\section{Discussion}\label{sec:Disc}
In this paper, we initiated general 2d $\cN=4$ superconformal bootstrap study, using the long-multiplets. As we have not specified any other properties of theory, other than $\cN=4$ superconformal symmetry, our analysis is general, but at the same time lack of decorations that could arise from global symmetries and analysis of BPS 4-point functions. This study provides the starting point for the numerical bootstrap analysis using the standard methods \cite{Simmons-Duffin:2016gjk,Simmons-Duffin:2015qma}. Also, since our superspace analysis is incomplete, it would be interesting to resolve the problem that we pointed out. Other than these obvious directions, there are several ways to use this set-up by imposing more input depending on the specific theories that preserve $\cN=4$ superconformal symmetry. 

Different from $\cN=2$ theories, $\cN=4$ theory has the stress energy tensor in short-multiplet. Rather than considering the long-multiplet 4-point function $\la\cL_0\cL_0\cL_0\cL_0\ra$, we can consider the short-multiplet 4-point function of $\cL_2$ that contains the stress energy tensor at the top. Because the stress energy tensor is a universal ingredient of any CFT \cite{Dymarsky:2017yzx}, this will also provide a general information on $\cN=4$ CFTs. Moreover, we expect a $\cL_2$ 4-point function, though it is BPS, may give a different restriction that $\cL_0$ 4-point function could not impose. Since the length of the multiplet and the number of components are reduced significantly in the BPS multiplet, we expect  efficient numerical analysis here.

CFTs with a global symmetry will give more stringent bounds, since there is a non-trivial relation between the level of Kac-Moody algebra and total central charge. Especially, there is a series of interesting $(0,4)$ theories with $E_8$ global symmetry that arises from IR limit of E-string worldsheet gauge theories \cite{Ganor:1996mu,Kim:2014dza}. The gauge theory lives on $N$ D2 brane worldvolume(012 direction); it has finite length(L) in direction 2 and extends between NS5 brane and D8/O8 complex. By taking $L$ small, there appears $2d$ $O(N)$ supersymmetric gauge theory with $SO(16)$ global symmetry. Flowing into deep IR(semi-classical limit or Higgs branch \cite{Witten:1997yu}), one expects to get $2d$ $(0,4)$ superconformal theory with a central charge $(c_L,c_R)=(6N,12N)$ and a global symmetry $E_8$. It would be interesting to study this series of CFT labeled by the number of E-strings and it would be also very interesting to see if there is another IR limit that comes from a different choice of IR R-symmetry, which was once suggested in \cite{Kim:2014dza}. Other big family of $(0,4)$ theories \cite{Putrov:2015jpa,Kapustin:2006hi} comes from a twisted compactification\footnote{We thank an anonymous referee of JHEP, who pointed out the original reference \cite{Kapustin:2006hi} for this topological twisting.} of class-$\cS$ theory, and \cite{Hanany:2018hlz} from the brane box model, which are another interesting models to study using the bootstrap technique.   

Lastly, our analysis can be used to study $4d$ $\cN=4$ SYM or SCFT, as 2d small $\cN=4$ chiral algebra appears in a particular twisted $Q-$cohomology of $4d$ $\cN=4$ SCFT \cite{Beem:2013sza}. \cite{Beem:2016wfs} mentioned this fact in their $4d$ $\cN=4$ numerical bootstrap analysis, but did an honest $4d$ superconformal block computation to construct 4-point functions and crossing equations. It would be interesting to use our result to study the $4d$ $\cN=4$ SCFT as we have much more crossing equations that can give more stringent bounds. 
\section*{Acknowledgements}
We thank Chi-Ming Chang for his collaboration in early stage of the project, especially his observation on the subtlety of $\cN=4$ superspace. We are also grateful to Ori Ganor for comments on the draft, and crucial advice in various stages of this project. We thank the organizers and participants in the 2017, 2018 Simons Bootstrap conference, where a part of the work was done. We especially thank Carlo Meneghelli for his comment on our paper, pointing out the possible resolution of our puzzle. This research was supported in part by the Berkeley Center of Theoretical Physics. The research of JO was supported in part by Kwanjeong Educational Foundation and by the Visiting Graduate Fellowship Program at the Perimeter Institute for Theoretical Physics. Research at the Perimeter Institute is supported by the Government of Canada through Industry Canada and by the Province of Ontario through the Ministry of Economic Development $\&$ Innovation.
\begin{appendix}
\section{3-point invariants of $\cN=4$ superspace}\label{app:integrand}
The idea is to start with arbitrary $3$ superspace coordinates
$(z_i,\theta_i,\btheta_i)$, with $i=1,2,3$, and perform superconformal transformations \footnote{We are grateful to Ori Ganor for sharing his unpublished notes that show preliminary result for the 3-point invariant of $\cN=4$ superspace \cite{Ori}} to set
$$
z_2=0,\quad
z_3=\infty,\quad
\theta_2=\theta_3=0,\quad
\btheta_2=\btheta_3=0,
$$
then, construct the dilatation invariant $\btheta_1'\theta_1'/z_1'$ from the resulting $(z_1',\theta_1',\btheta_1')$. The details are below.

We use following conventions (for $\alpha,\beta=1,2$ and $a=1,2,3$):
\ie
\theta_\alpha=\epsilon_{\alpha\beta}\theta^\beta,
~~\btheta^\alpha=\epsilon^{\alpha\beta}\btheta_\beta,
~~\sigma^a:=(\sigma^a)^\alpha{}_\beta,
~~(\sigma^i)_\alpha{}^\beta=\epsilon_{\alpha\alpha'}\epsilon^{\beta\beta'}(\sigma^a)^{\alpha'}{}_{\beta'},
~~\epsilon^{\alpha\beta}=-\epsilon_{\alpha\beta}
\fe
For two doublets $\psi^\alpha$ and $\chi^\alpha$,
$$
\psi\chi:=\psi^\alpha\chi_\alpha = \epsilon_{\alpha\beta}\psi^\alpha\chi^\beta
=-\chi_\alpha\psi^\alpha = \chi^\alpha\psi_\alpha=\chi\psi
$$

Inversion acts on the superspace coordinates by
\ie
\mathcal{I}:~(z,\theta,\bar{\theta})~\rightarrow~(-1/z,\theta/z,\btheta/z)
\fe
Also, as usual, rigid SUSY with parameters $\pSUSY^\alpha$ acts as
\be\label{eqn:rigidSUSY}
\delta z_i = -\bpSUSY\theta_i+\pSUSY\btheta_i
\,,\qquad
\delta\theta_i = \pSUSY
\,,\qquad
\delta\btheta_i = \bpSUSY
\ee
Denote
$$
z_{ij}:= z_i - z_j,
\qquad
\theta_{ij}:=\theta_i-\theta_j,
\qquad
\btheta_{ij}:=\btheta_i-\btheta_j.
$$
Then $\theta_{ij}$, $\btheta_{ij}$, and
$$
\sZ_{ij}:=z_{ij}+\theta_i\btheta_j-\theta_j\btheta_i
$$
are invariant under \eqref{eqn:rigidSUSY}.

A noninfinitesimal SUSY transformation with parameters $\eta$ and $\oeta$ acts as
$$
z\rightarrow z-\oeta\theta+\eta\btheta
\,,\qquad
\theta\rightarrow\theta+\eta
\,,\qquad
\btheta\rightarrow\btheta+\oeta
$$
Then we construct a large superconformal transformation from a translation by $(\zeta_1,\eta_1,\oeta_1)$ followed by inversion, followed by translation by $(\zeta_2,\eta_2,\oeta_2)$, followed by dilatation by $\lambda$ (the dilatation will be implicit).
$$
z\rightarrow z-\oeta_1\theta+\eta_1\btheta+\zeta_1
\rightarrow -\frac{1}{z-\oeta_1\theta+\eta_1\btheta+\zeta_1}
\,,~
\theta\rightarrow \frac{\theta+\eta_1}{z-\oeta_1\theta+\eta_1\btheta+\zeta_1}
\,,~
\btheta\rightarrow\frac{\btheta+\oeta_1}{z-\oeta_1\theta+\eta_1\btheta+\zeta_1}
$$
Next,
$$
\frac{\theta+\eta_1}{z-\oeta_1\theta+\eta_1\btheta+\zeta_1}
\rightarrow
\frac{\theta+\eta_1}{z-\oeta_1\theta+\eta_1\btheta+\zeta_1}+\eta_2
\,,~
\frac{\btheta+\oeta_1}{z-\oeta_1\theta+\eta_1\btheta+\zeta_1}
\rightarrow
\frac{\btheta+\oeta_1}{z-\oeta_1\theta+\eta_1\btheta+\zeta_1}+\oeta_2\,,
$$
\bear
\lefteqn{
-\frac{1}{z-\oeta_1\theta+\eta_1\btheta+\zeta_1}
\rightarrow
}\nn\\ &&
-\frac{1}{z-\oeta_1\theta+\eta_1\btheta+\zeta_1}
-\oeta_2\left(
\frac{\theta+\eta_1}{z-\oeta_1\theta+\eta_1\btheta+\zeta_1}
\right)
+\eta_2\left(
\frac{\btheta+\oeta_1}{z-\oeta_1\theta+\eta_1\btheta+\zeta_1}
\right)+\zeta_2
\nn
\eear
So, altogether
$$
z\rightarrow z':=
\zeta_2
+\frac{-\oeta_2\theta+\eta_2\btheta-\oeta_2\eta_1+\oeta_1\eta_2-1}{z-\oeta_1\theta+\eta_1\btheta+\zeta_1}
$$
$$
\theta\rightarrow\theta':=
\frac{\theta+\eta_1}{z-\oeta_1\theta+\eta_1\btheta+\zeta_1}+\eta_2
\,,\qquad
\btheta\rightarrow\btheta':=
\frac{\btheta+\oeta_1}{z-\oeta_1\theta+\eta_1\btheta+\zeta_1}+\oeta_2
$$
Now we start with three superspace coordinates $(z_i,\theta_i,\btheta_i)$ with $i=1,2,3$.
Let us first set $z'_3=\infty$ by setting
$$
\zeta_1=-z_3+\oeta_1\theta_3-\eta_1\btheta_3\,.
$$
Next, we require $\theta_3'$ and $\btheta_3'$ to be finite (and therefore zero after inversion) by setting
$$
\eta_1=-\theta_3,\qquad
\oeta_1=-\btheta_3
$$
Thus,
$$
\zeta_1=-z_3+\oeta_1\theta_3-\eta_1\btheta_3
=-z_3-\btheta_3\theta_3+\theta_3\btheta_3
=-z_3
$$
Next, we require $z'_2=0$ by setting
$$
\zeta_2 = 
\frac{\oeta_2\theta_2-\eta_2\btheta_2+\oeta_2\eta_1-\oeta_1\eta_2+1}{z_2-\oeta_1\theta_2+\eta_1\btheta_2+\zeta_1}
=
\frac{\oeta_2\theta_2-\eta_2\btheta_2-\oeta_2\theta_3+\btheta_3\eta_2+1}{z_{23}+\btheta_3\theta_2-\theta_3\btheta_2} =\frac{\oeta_2\theta_{23}-\eta_2\btheta_{23}+1}{Z_{23}}
$$
We also require $\theta'_2=\btheta'_2=0$ by setting
$$
0=
\frac{\theta_2+\eta_1}{z_2-\oeta_1\theta_2+\eta_1\btheta_2+\zeta_1}+\eta_2
$$
and
$$
0=\frac{\btheta_2+\oeta_1}{z_2-\oeta_1\theta_2+\eta_1\btheta_2+\zeta_1}+\oeta_2
$$
Thus
\be\label{eqn:eta2}
\eta_2 = -\frac{\theta_2+\eta_1}{z_2-\oeta_1\theta_2+\eta_1\btheta_2+\zeta_1}
=-\frac{\theta_{23}}{Z_{23}}
\ee
\be\label{eqn:oeta2}
\oeta_2 = -\frac{\btheta_2+\oeta_1}{z_2-\oeta_1\theta_2+\eta_1\btheta_2+\zeta_1}
=-\frac{\btheta_{23}}{Z_{23}}
\ee

After this transformation, we are left with
\bear
\lefteqn{
z_1'=
\zeta_2
+\frac{-\oeta_2\theta_1+\eta_2\btheta_1-\oeta_2\eta_1+\oeta_1\eta_2-1}{z_1-\oeta_1\theta_1+\eta_1\btheta_1+\zeta_1}
}\nn\\
&=&
\frac{\oeta_2\theta_2-\eta_2\btheta_2-\oeta_2\theta_3+\btheta_3\eta_2+1}{Z_{23}}
+\frac{-\oeta_2\theta_1+\eta_2\btheta_1+\oeta_2\theta_3-\btheta_3\eta_2-1}{Z_{13}}
\nn\\
&=&
\frac{\oeta_2\theta_{23}-\eta_2\btheta_{23}+1}{Z_{23}}
+\frac{-1-\oeta_2\theta_{13}+\eta_2\btheta_{13}}{Z_{13}}\\
&=&
\frac{Z_{13}-Z_{23}+\theta_{13}\bar{\theta}_{23}-\theta_{23}\bar{\theta}_{13}}{Z_{23}Z_{13}}=\frac{Z_{12}}{Z_{13}Z_{23}}
\label{eqn:z1p}
\eear
with $\eta_2$ and $\oeta_2$ as above,
and
\be\label{eqn:th1p}
\theta_1' =
\frac{\theta_1+\eta_1}{z_1+\oeta_1\theta_1-\eta_1\btheta_1+\zeta_1}+\eta_2
=
\frac{\theta_{13}}{Z_{13}}
-\frac{\theta_{23}}{Z_{23}}
\ee
\be\label{eqn:bth1p}
\btheta_1' =
\frac{\btheta_1+\oeta_1}{z_1+\oeta_1\theta_1-\eta_1\btheta_1+\zeta_1}+\oeta_2
=
\frac{\btheta_{13}}{Z_{13}}
-\frac{\btheta_{23}}{Z_{23}}
\ee
We still have dilatation freedom and $SU(2)_R$ freedom, and if we also require $U(1)$ invariance, we are left with one overall invariant
\be\label{eqn:U123}
U_{123}:=\frac{\btheta_1'\theta_1'}{z_1'}
=\cdots
\ee
We have to substitute \eqref{eqn:eta2}, \eqref{eqn:oeta2}, \eqref{eqn:z1p}, \eqref{eqn:th1p}, \eqref{eqn:bth1p} into \eqref{eqn:U123} to get the full conformal invariant.
But we can check what $U_{123}$ looks like at $O(\theta^2)$.
We have
$$
z_1' = \frac{1}{z_{13}}-\frac{1}{z_{23}}+O(\theta^2) = -\frac{z_{12}}{z_{13}z_{23}}+O(\theta^2)
$$
and
$$
\btheta_1'=\frac{z_{23}\btheta_{13}-z_{13}\btheta_{23}}{z_{13}z_{23}}
\,,\qquad
\theta_1'=\frac{z_{23}\theta_{13}-z_{13}\theta_{23}}{z_{13}z_{23}}
$$
So,
$$
U_{123} = 
\frac{z_{23}\btheta_{13}\theta_{13}}{z_{12}z_{13}}
-\frac{\btheta_{23}\theta_{13}+\btheta_{13}\theta_{23}}{z_{12}}
+\frac{z_{13}\btheta_{23}\theta_{23}}{z_{12}z_{23}}+O(\theta^4)
$$
We can write
$$
\frac{z_{23}}{z_{12}z_{13}}=\frac{1}{z_{12}}-\frac{1}{z_{13}}
\,,\qquad
\frac{z_{13}}{z_{12}z_{23}}=\frac{1}{z_{12}}+\frac{1}{z_{23}}
$$
to simplify the above expression.

More explicitly, the $U_{123}$ is
\ie
U_{123}&=\frac{\btheta_1'\theta_1'}{z_1'}=\frac{Z_{13}Z_{23}}{Z_{12}}\left(\frac{\theta_{13}}{Z_{13}}
-\frac{\theta_{23}}{Z_{23}}\right)\left(\frac{\bar\theta_{13}}{Z_{13}}
-\frac{\bar\theta_{23}}{Z_{23}}\right)
\\
&={(\theta_{13}Z_{23}-\theta_{23}Z_{13})(\bar\theta_{13}Z_{23}-\bar\theta_{23}Z_{13})\over Z_{12}Z_{13}Z_{23}}
\\
&={\theta_{13}\bar\theta_{13}Z_{23}^2+\theta_{23}\bar\theta_{23}Z_{13}^2-Z_{13}Z_{23}(\theta_{13}\bar\theta_{23}+\theta_{23}\bar\theta_{13})\over Z_{12}Z_{13}Z_{23}}.
\fe
\section{2-point function normalization}\label{2ptnorm}
Here, we collected all relevant 2-point function normalization. We also submitted Mathematica files that have the same information.
\subsection{$\cL_0$}
We order and number each component fields of $\cL_0$ from bottom component to top component.
\ie\label{l0set}
\cL_0=\{\phi,\psi^1,\psi^2,\chi^1,\chi^2,\tau,\bar{\tau},t_0,t_t^1,t_t^2,t_t^3,C^1,C^2,\bar{C}^1,\bar{C}^2,d\}=\{\cF_1,\cF_2,\ldots,\cF_{16}\}
\fe
Below, $f_{i,j}$ refers to $\la\cF_i\cF_j\ra$ normalization constant.
\ie
&f_{1,1}=F_0,~f_{2,5}=f_{5,2}=2hF_0,~f_{3,4}=f_{4,3}=2hF_0,~f_{6,7}=f_{7,6}=4h(1+h)F_0,~f_{8,8}=8h(1+h)F\\
&f_{9,11}=f_{11,9}=4h^2F_0,~f_{10,10}=2h^2F_0,~f_{12,15}=f_{15,12}=\frac{16h^2(1+h)^2}{1+2h}F_0,\\
&f_{13,14}=f_{14,13}=\frac{16h^2(1+h)^2}{1+2h}F_0,~f_{16,16}=\frac{16h^2(1+h)^2(3+2h)}{1+2h}F_0
\fe
In other words, all the normalization constants are determined up to a constant $F_0$.
\subsection{$\cL_1$}
We first fix the order and number the components
\ie\label{l1set}
\cL_1=&\{\phi[1],\psi[0],\psi[2],\chi[0],\chi[2],\tau[1],\bar{\tau}[1],t_1[1],t_2[1],t[3],C[0],C[2],\bar{C}[0],\bar{C}[2],d[1]\}=\{\cF_1,\cF_2,\ldots,\cF_{32}\}
\fe
Here, for non-trivial representations of $su(2)_R$, such as $\phi[1]$, we aligned from bottom component to top component of R-symmetry multiplet. For instance, $\phi[1]=\{\cF_1,\cF_2\}$, $\psi[0]=\{\cF_3\}$, $\psi[2]=\{\cF_4,\cF_5,\cF_6\}$. Below, $f_{i,j}$ refers to $\la\cF_i\cF_j\ra$ normalization constant.\footnote{In practical use in numerics, one needs positive normalization. Since the - signs in some of $f_{i,j}$ come from our definitions of operators, they further need to be re-defined.}
\ie\nonumber
&f_{1,2}=-f_{2,1}=F_1,~f_{3,7}=f_{7,3}=-2(3+2h)f_{1,2},~f_{4,10}=-f_{10,4}=(1-2h)F_1,~f_{5,9}=-f_{9,5}=\frac{1-2h}{2}F_1,\\
&f_{6,8}=f_{8,6}=(1-2h)F_1,~f_{11,14}=-f_{14,11}=(1-2h)(3+2h)F_1,~f_{12,13}=-f_{13,12}=(2h-1)(3+2h)F_1,
\fe
\ie
&f_{15,16}=-f_{16,15}=\frac{(1-2h)(3+2h)}{2h}F_1,~f_{15,18}=-f_{18,15}=\frac{(1-2h)(3+2h)(3+4h)}{2h}F_1,\\
&f_{16,17}=-f_{17,16}=\frac{(2h-1)(3+2h)(3+4h)}{2h}F_1,~f_{17,18}=-f_{18,17}=\frac{(1-2h)(3+2h)^2}{2h}F_1,\\
&f_{19,22}=-f_{22,19}=-(1-2h)^2F_1,~f_{20,21}=\frac{(1-2h)^2}{3}F_1,~f_{23,27}=-f_{27,23}=\frac{4(1+h)(1-2h)(3+2h)^2}{1+2h}F_1,\\
&f_{24,30}=-f_{30,27}=-\frac{2(1-2h)^2(1+h)(3+2h)}{1+2h}F_1,~f_{25,29}=-f_{29,25}=\frac{(1-2h)^2(1+h)(3+2h)}{1+2h}F_1,\\
&f_{31,32}=-f_{32,31}=\frac{(1-2h)^2(3+2h)^3}{1+2h}F_1
\fe
Similar to above, all the 2-point normalizations are fixed up to a constant $F_1$.
\subsection{$\cL_2$}
We first fix the order and number the components
\ie\label{l2set}
\cL_2=&\{\phi[2],\psi[1],\psi[3],\chi[1],\chi[3],\tau[2],\bar{\tau}[2],t[0],t_1[2],t_2[2],t[4],C[1],C[3],\bar{C}[1],\bar{C}[3],d[2]\}\\
=&\{\cF_1,\ldots,\cF_{48}\}
\fe
Similarly, we pick the same order in the R-symmetry multiplet as above.
\ie\nonumber
&f_{1,3}=F_2,~f_{2,2}=\frac{1}{2}F_2,~f_{4,11}=f_{11,4}=3(2+h)F_2,~f_{5,10}=f_{10,5}=3(2+h)F_2,~\\
&f_{6,15}=f_{15,6}=2(h-1)F_2~f_{7,14}=f_{14,7}=\frac{2(-1+h)}{3}F_2,~f_{8,13}=f_{13,8}=\frac{2(1-h)}{3}F_2,\\
&f_{9,12}=f_{12,9}=2(-1+h)F_2,~f_{16,21}=f_{21,16}=4(1-h)(2+h)F_2,~\\
&f_{17,20}=f_{20,17}=2(h-1)(2+h)F,~f_{18,19}=f_{19,18}=4(1-h)(2+h)F_2,~f_{22,22}=12(2+h)^2F_2,\\
&~f_{23,25}=f_{25,23}=\frac{2(1-h)(2+h)^2}{h}F_2,~f_{23,28}=\frac{2(1-h)(4+8h+3h^2)}{h}F_2,\\
&f_{24,24}=\frac{(h-1)(2+h)^2}{h}F_2,~f_{24,27}=f_{27,24}=\frac{(-1+h)(2+h)(2+3h)}{h}F_2,\\
&f_{25,26}=f_{26,25}=\frac{2(1-h)(4+8h+3h^2)}{h}F_2,~f_{26,28}=f_{28,26}=\frac{2(1-h)(2+h)^2}{h}F_2,\\
&f_{29,33}=4(h-1)^2F_2,~f_{30,32}=f_{32,30}=(h-1)^2F_2,~f_{31,31}=\frac{2}{3}(h-1)^2F,\\
&f_{34,41}=f_{41,34}=\frac{24(1-h)(1+h)(2+h)^2}{1+2h}F_2,~f_{35,40}=f_{40,35}=\frac{24(h-1)(1+h)(2+h)^2}{1+2h}F_2,
\fe
\ie
&f_{36,45}=f_{45,36}=\frac{16(h-1)^2(2+3h+h^2)}{1+2h}F_2,~f_{37,44}=f_{44,37}=\frac{16(h-1)^2(2+3h+h^2)}{3+6h}F_2,\\
&f_{38,43}=f_{43,38}=\frac{16(h-1)^2(2+3h+h^2)}{3+6h}F_2,~f_{39,42}=f_{42,39}=\frac{16(-1+h)^2(2+3h+h^2)}{1+2h}F_2,\\
&f_{46,48}=f_{48,46}=\frac{16(3+2h)(-2+h+h^2)^2}{1+2h}F_2,~f_{47,47}=\frac{8(3+2h)(-2+h+h^2)^2}{1+2h}F_2
\fe
\section{3-point function normalizations}\label{3ptnorm}
Since there are too many 3-point functions, here we only present those of $\la\phi\cL_0\cL_0\ra$, $\la\phi\cL_0\cL_1\ra$, $\la\phi\cL_0\cL_2\ra$ with $\phi\in\cL_0$. $\cL_0$, $\cL_1$, $\cL_2$ are built from superconformal primary $\phi$, $\phi^\A$, $\phi^{\A\B}$ with weight $h$. The most general results with different weights and $\la\cL_0\cL_0\cL_0\ra$, $\la\cL_0\cL_0\cL_1\ra$, $\la\cL_0\cL_0\cL_2\ra$ can be found in the separate Mathematica file that we submitted.

Let us describe how to read off the result from our mathematica file. The script consists of the substitution rules for all 3-point coefficients. As we described, they are expressed fully in terms of 10 independent constants. $f[i,j,k]$ indicates 3-point function OPE coefficients of three fields $\cF_i$, $\cF_j$, $\cF_k$, where the $i,j$ indices run in \eqref{l0set}, and $k$ index runs in \eqref{l0set},\eqref{l1set},\eqref{l2set} in each file. $h[1]$, $h[2]$, $h[3]$ is the conformal weight of superconformal primary of $\cF_i$, $\cF_j$, $\cF_k$, respectively.
\subsection{$\la\phi\cL_0\cL_0\ra$}
\ie\nonumber
   &f_{1,1,1}= f_{1,1,1},~f_{1,1,6}= f_{1,1,6},~f_{1,1,7}= f_{1,1,7},~f_{1,1,8}=
   f_{1,1,8},~f_{1,1,16}= f_{1,1,16},~f_{1,2,3}= f_{1,1,6},\\
   &f_{1,2,4}= -hf_{1,1,1}-f_{1,1,7},~f_{1,2,13}= h (h+1) f_{1,1,6},~f_{1,2,14}= \frac{1}{2}
   \left(-2 f_{1,1,7} h^2-2 f_{1,1,7} h+f_{1,1,16}\right),\\
   &f_{1,3,2}=-f_{1,1,6},~f_{1,3,5}= -h f_{1,1,1}-f_{1,1,7},~f_{1,3,12}= -h (h+1)
   f_{1,1,6},\\
   &f_{1,3,15}= \frac{1}{2} \left(-2 f_{1,1,7} h^2-2 f_{1,1,7}
   h+f_{1,1,16}\right),~f_{1,4,2}= f_{1,1,7}-h f_{1,1,1},~f_{1,4,5}= f_{1,1,8},\\
   &f_{1,4,12}= f_{1,1,7} h^2+f_{1,1,7} h+\frac{1}{2},~f_{1,1,16},~f_{1,4,15}= h (h+1) f_{1,1,8},~f_{1,5,3}= f_{1,1,7}-h
   f_{1,1,1},~f_{1,5,4}= -f_{1,1,8},\\
   &f_{1,5,13}= f_{1,1,7} h^2+f_{1,1,7} h+\frac{1}{2}f_{1,1,16},~f_{1,5,14}= -h (h+1) f_{1,1,8},~f_{1,6,1}= f_{1,1,6},~f_{1,6,7}=-(h+1) f_{1,1,6},\\
      &f_{1,6,8}= \frac{-2 f_{1,1,1} h^3-4
   \left(f_{1,1,1}+f_{1,1,7}\right) h^2-2 \left(f_{1,1,1}+3 f_{1,1,7}\right) h-2
   f_{1,1,7}+f_{1,1,16}}{2 h+1},\\
   &f_{1,6,16}= -2 h \left(h^2+3 h+2\right)
   f_{1,1,6},~f_{1,7,1}= f_{1,1,7},~f_{1,7,6}= (h+1) f_{1,1,6},\\
   &f_{1,7,7}= \frac{2
   f_{1,1,1} h^3+4 f_{1,1,1} h^2+2 f_{1,1,1} h-f_{1,1,16}}{4 h+2},~f_{1,7,8}= -(h+1)
   f_{1,1,8},\\
   &f_{1,7,16}= -2 h \left(h^2+3 h+2\right) f_{1,1,7},~f_{1,8,1}=
   f_{1,1,8},\\
   &f_{1,8,6}= \frac{-2 f_{1,1,1} h^3-4 \left(f_{1,1,1}-f_{1,1,7}\right)
   h^2-2 \left(f_{1,1,1}-3 f_{1,1,7}\right) h+2 f_{1,1,7}+f_{1,1,16}}{2
   h+1},\\
      &f_{1,8,7}= (h+1) f_{1,1,8},~f_{1,8,16}= -2 h \left(h^2+3 h+2\right)
   f_{1,1,8},~f_{1,10,10}= -\frac{2 f_{1,1,1} h^3+2 f_{1,1,1} h^2+f_{1,1,16}}{4
   h+2},\\
   &f_{1,11,11}= \frac{2 f_{1,1,1} h^3+2 f_{1,1,1} h^2+f_{1,1,16}}{4
   h+2},~f_{1,12,3}= -h (h+1) f_{1,1,6},~f_{1,12,4}= f_{1,1,7} h^2+f_{1,1,7}
   h-\frac{1}{2} f_{1,1,16},\\
   &f_{1,12,13}= -h^2 \left(h^2+3 h+2\right)
   f_{1,1,6},\\  
   &f_{1,12,14}= \frac{1}{2} (h+2) \left(2 f_{1,1,1} h^4+2 \left(2
   f_{1,1,1}+f_{1,1,7}\right) h^3+2 \left(f_{1,1,1}+f_{1,1,7}\right)
   h^2+f_{1,1,16}\right),\\
   &f_{1,13,2}= h (h+1) f_{1,1,6},~f_{1,13,5}= f_{1,1,7}
   h^2+f_{1,1,7} h-\frac{1}{2} f_{1,1,16},~f_{1,13,12}= h^2 \left(h^2+3 h+2\right)
   f_{1,1,6},
   \fe
   \ie
   &f_{1,13,15}= \frac{1}{2} (h+2) \left(2 f_{1,1,1} h^4+2 \left(2
   f_{1,1,1}+f_{1,1,7}\right) h^3+2 \left(f_{1,1,1}+f_{1,1,7}\right)
   h^2+f_{1,1,16}\right),\\
   &f_{1,14,2}= -f_{1,1,7} h^2-f_{1,1,7} h-\frac{1}{2}
   f_{1,1,16},~f_{1,14,5}= -h (h+1) f_{1,1,8},\\
   &~f_{1,14,12}= \frac{1}{2} (h+2) \left(2
   f_{1,1,1} h^4+\left(4 f_{1,1,1}-2 f_{1,1,7}\right) h^3+2
   \left(f_{1,1,1}-f_{1,1,7}\right) h^2+f_{1,1,16}\right),\\
   &f_{1,14,15}= -h^2\left(h^2+3 h+2\right) f_{1,1,8},~f_{1,15,3}= -f_{1,1,7} h^2-f_{1,1,7} h-\frac{1}{2}
   f_{1,1,16},~f_{1,15,4}= h (h+1) f_{1,1,8},\\
   &f_{1,15,13}= \frac{1}{2} (h+2) \left(2
   f_{1,1,1} h^4+\left(4 f_{1,1,1}-2 f_{1,1,7}\right) h^3+2
   \left(f_{1,1,1}-f_{1,1,7}\right) h^2+f_{1,1,16}\right),\\
   &f_{1,15,14}= h^2 \left(h^2+3
   h+2\right) f_{1,1,8},~f_{1,16,1}= f_{1,1,16},~f_{1,16,6}= -2 h \left(h^2+3
   h+2\right) f_{1,1,6},\\
   &f_{1,16,7}= -2 h \left(h^2+3 h+2\right)
   f_{1,1,7},~f_{1,16,8}= -2 h \left(h^2+3 h+2\right) f_{1,1,8},\\
   &f_{1,16,16}= 2
   \left(h^2+5 h+6\right) \left(2 f_{1,1,1} h^4+4 f_{1,1,1} h^3+2 f_{1,1,1}
   h^2+f_{1,1,16}\right) 
   \fe
   \subsection{$\la\phi\cL_0\cL_1\ra$}
   \begin{center}
   \ie\nonumber
   &f_{1,1=,3}= f_{1,1,3},~f_{1,1,7}= f_{1,1,7},~f_{1,1,23}=
   f_{1,1,23},~f_{1,1,27}= f_{1,1,27},~f_{1,2,2}= \frac{1}{2} f_{1,1,3},\\
   &f_{1,2,14}=
   \frac{2 (2 h+1) f_{1,1,27}-(-2 h-1) (2 h-1) f_{1,1,7}}{8 h+4},~f_{1,2,16}=
   f_{1,1,23}-\frac{(-2 h-1) \left(4 h^2+4 h-3\right) f_{1,1,3}}{8 h (2
   h+1)},\\
   &f_{1,2,18}= \frac{1}{2} f_{1,1,23}-\frac{(-2 h-1) \left(8 h^2+2 h-3\right)
   f_{1,1,3}}{8 h (2 h+1)},~f_{1,2,32}= -\frac{(-2 h-3) (2 h-1) f_{1,1,23}}{8
   (h+1)},\\
   &f_{1,4,2}= \frac{1}{2} f_{1,1,7},~f_{1,4,12}= \frac{(-2 h-1) (2 h-1)
   f_{1,1,3}+2 (2 h+1) f_{1,1,23}}{8 h+4},\\
   &f_{1,4,16}= -\frac{(-2 h-1) \left(8 h^2+2
   h-3\right) f_{1,1,7}}{8 h (2 h+1)}-\frac{1}{2} f_{1,1,27},~f_{1,4,18}= -\frac{(-2
   h-1) \left(4 h^2+4 h-3\right) f_{1,1,7}}{8 h (2 h+1)}-f_{1,1,27},\\
   &f_{1,4,32}=
   \frac{(-2 h-3) (2 h-1) f_{1,1,27}}{8 (h+1)},~f_{1,6,7}= \frac{2 (2 h+1)
   f_{1,1,23}-(-2 h-1) (2 h+3) f_{1,1,3}}{4 h+2},\\
   &f_{1,6,27}= -\frac{(-2 h-3) (2 h-1)
   \left((-2 h-1) (2 h+3) f_{1,1,3}-2 (2 h+1) f_{1,1,23}\right)}{4 (2
   h+1)^2},\\
   &f_{1,7,3}= \frac{(-2 h-1) (2 h+3) f_{1,1,7}+2 (2 h+1) f_{1,1,27}}{4
   h+2},\\
   &f_{1,7,23}= -\frac{(-2 h-3) (2 h-1) \left((-2 h-1) (2 h+3) f_{1,1,7}+2 (2 h+1)
   f_{1,1,27}\right)}{4 (2 h+1)^2},\\
   &f_{1,8,3}= \frac{2 (2 h+1) f_{1,1,23}-(-2 h-1) (2
   h+3) f_{1,1,3}}{4 h+2},~f_{1,8,7}= \frac{(-2 h-1) (2 h+3) f_{1,1,7}+2 (2 h+1)
   f_{1,1,27}}{4 h+2},\\
   &f_{1,8,23}= \frac{(-2 h-3) (2 h-1) \left((-2 h-1) (2 h+3)
   f_{1,1,3}-2 (2 h+1) f_{1,1,23}\right)}{4 (2 h+1)^2},\\
   &f_{1,8,27}= \frac{(-2 h-3) (2
   h-1) \left((-2 h-1) (2 h+3) f_{1,1,7}+2 (2 h+1) f_{1,1,27}\right)}{4 (2
   h+1)^2},\\
   &f_{1,9,6}= -\frac{(-2 h-1) (2 h-1) f_{1,1,3}+2 (2 h+1) f_{1,1,23}}{8
   h+4},~f_{1,9,10}= \frac{2 (2 h+1) f_{1,1,27}-(-2 h-1) (2 h-1) f_{1,1,7}}{8
   h+4},
      \fe
   \ie
   &f_{1,9,26}= -\frac{(-2 h-3) (2 h-1) \left((-2 h-1) (2 h-1) f_{1,1,3}+2 (2 h+1)
   f_{1,1,23}\right)}{8 (2 h+1)^2},\\
         &f_{1,9,30}= \frac{(-2 h-3) (2 h-1) \left((-2 h-1)
   (2 h-1) f_{1,1,7}-2 (2 h+1) f_{1,1,27}\right)}{8 (2 h+1)^2},\\
      &f_{1,12,2}= \frac{1}{4}
   \left(f_{1,1,3}+2 f_{1,1,23}\right),~f_{1,12,14}= -\frac{(h+1) (2 h+3) \left(2 (2
   h+1) f_{1,1,27}-(-2 h-1) (2 h-1) f_{1,1,7}\right)}{4 (2 h+1)^2},\\
      &f_{1,12,16}=
   \frac{(-2 h-3) \left(-(2 h+1) (4 h+1) \left(4 h^2+4 h-3\right) f_{1,1,3}-2 (2 h+1)
   ((4 h-6) h-2 h-3) f_{1,1,23}\right)}{16 h (2 h+1)^2},\\
   &f_{1,12,18}= \frac{(-2 h-3)
   \left((-2 h-1) (2 h-1) (4 (h+3) h-2 h+3) f_{1,1,3}-2 (2 h+1) ((8 h-6) h+2 h-3)
   f_{1,1,23}\right)}{16 h (2 h+1)^2},\\
        &f_{1,12,32}= -\frac{(2 h-1) \left(12 h^2+8 (h+2)
   h-8 (2 h+2) h+16 h+15\right) (2 (2 h+1) (6 h+3) f_{1,1,23}}{32 (h+1) (2 h+1)^3},\\
   &\frac{-(2 h+1)^2 \left(4
   h^2+4 h-3\right) f_{1,1,3})}{32 (h+1) (2 h+1)^3}\\
   &f_{1,14,2}= \frac{1}{4}
   \left(2 f_{1,1,27}-f_{1,1,7}\right),~f_{1,14,12}= \frac{(h+1) (2 h+3) \left((-2 h-1)
   (2 h-1) f_{1,1,3}+2 (2 h+1) f_{1,1,23}\right)}{4 (2 h+1)^2},\\
   &f_{1,14,16}= -\frac{(-2
   h-3) \left((-2 h-1) (2 h-1) (4 (h+3) h-2 h+3) f_{1,1,7}+2 (2 h+1) ((8 h-6) h+2 h-3)
   f_{1,1,27}\right)}{16 h (2 h+1)^2},\\
   &f_{1,14,18}= -\frac{(-2 h-3) \left(2 (2 h+1) ((4
   h-6) h-2 h-3) f_{1,1,27}-(2 h+1) (4 h+1) \left(4 h^2+4 h-3\right)
   f_{1,1,7}\right)}{16 h (2 h+1)^2},\\
   &f_{1,14,32}= -\frac{(2 h-1) \left(12 h^2+8 (h+2)
   h-8 (2 h+2) h+16 h+15\right)}{32 (h+1) (2 h+1)^3}\\
   &\times  \left(\left(4 h^2+4 h-3\right) f_{1,1,7} (2 h+1)^2+2 (6
   h+3) f_{1,1,27} (2 h+1)\right)
   &f_{1,16,3}= -\frac{(2 h+3)
   \left(2 (2 h+1)^2 f_{1,1,23}-(-2 h-1) (2 h+1) f_{1,1,3}\right)}{4 (2
   h+1)^2},\\
   &f_{1,16,7}= -\frac{(-2 h-3) \left(2 f_{1,1,27} (2 h+1)^2+(-2 h-1) f_{1,1,7}
   (2 h+1)\right)}{4 (2 h+1)^2},\\
   &f_{1,16,23}= \frac{ \left(2 (2 h+1) (6 h+3) f_{1,1,23}-(2 h+1)^2 \left(4 h^2+4
   h-3\right) f_{1,1,3}\right)}{8 (2 h+1)^3}\\
   &\times \left(12 h^2+8 (h+2) h-8 (2 h+2)
   h+16 h+15\right)\\
   &f_{1,16,27}= \frac{-2 (-2 h-5) h (2 h-1)
   (2 h+3)^2 f_{1,1,7} (2 h+1)^3}{16 h (2 h+1)^4}\\
   &+\frac{-\left(16 h^4-16 (4 h+3) h^3-16 h^2+4 (4
   h+3)^3 h-(4 h+3)^2 \left(8 h^2+4 (2 h+3) h+12 h+5\right)\right) f_{1,1,27} (2
   h+1)^2}{(2 h+1)^4(8 h+4)}
   \fe
   \end{center}
   \subsection{$\la\phi\cL_0\cL_2\ra$}
   \ie\nonumber
   &f_{1,1,22}= f_{1,1,22},~f_{1,2,11}= \frac{1}{2} f_{1,1,22},~f_{1,2,41}=
   \frac{(-h-1) (h-1) f_{1,1,22}}{2 h+1},~f_{1,4,5}= -\frac{1}{2}
   f_{1,1,22},\\
   &f_{1,4,35}= \frac{(-h-1) (h-1) f_{1,1,22}}{2 h+1},~f_{1,9,3}=
   \frac{1}{3} f_{1,1,22},~f_{1,9,25}= -\frac{2 (-h-1) (h-1) f_{1,1,22}}{3
   h},\\
   &f_{1,9,28}= -\frac{2 (-h-1) (h-1) f_{1,1,22}}{3 h},~f_{1,9,48}= \frac{2 (-h-2)
   (-h-1) (h-1)^2 f_{1,1,22}}{3 (h+1) (2 h+1)},
   \fe
   \ie
   &f_{1,12,11}= -\frac{(h+1)^2
   f_{1,1,22}}{2 h+1},\\
   &f_{1,12,41}= \frac{2 (h-1) (h+1) \left(3 h^2+(2 h+3) h-(4 h+3)
   h+3 h+2\right) f_{1,1,22}}{(2 h+1)^2},~f_{1,14,5}= -\frac{(h+1)^2 f_{1,1,22}}{2
   h+1},\\
   &f_{1,14,35}= -\frac{2 (h-1) (h+1) \left(3 h^2+(2 h+3) h-(4 h+3) h+3 h+2\right)
   f_{1,1,22}}{(2 h+1)^2},\\
   &f_{1,16,22}= \frac{2 (h+1) \left(3 h^2+(2 h+3) h-(4 h+3) h+3
   h+2\right) f_{1,1,22}}{2 h+1}
   \fe
   \section{Sample crossing equations}\label{app:crossing equations}
   In this appendix, we collect sample crossing equations obtained from $\la\psi^1\chi^2\phi\phi\ra$, $\la\phi\phi\phi\phi\ra$ and $\la\psi^2C^2\phi\phi\ra$. We will use following notation for Virasoro conformal block:
   \ie
   g[z,\bar{z},h_{12},h_{34},h_{ex}-h]&=g_{h_{ex}}^{h_{12},h_{34}}(z)g_{\bar{h}_{ex}}^{\bar{h}_{12},\bar{h}_{34}}(\bar{z})\\
   g_{h_{ex}}^{h_{12},h_{34}}(z)&=z^{h_{ex}}{}_2F_1(h_{ex}-h_{12},h_{ex}+h_{34},2h_{ex},z)
   \fe
   where $h$ is the conformal weight for superconformal primary of $\cL_0,\cL_1,\cL_2$, and $h_{ex}$ is the conformal weight for exchanged operator.
   
   In the mathematica file, we showed all the crossing equations that were obtained by studying long-multiplet 4-point function of $\cL_0$. $H$ is the conformal weight of the superconformal primary of $\cL_0$.
   
   Crossing equations obtained from $\la\psi^1\chi^2\phi\phi\ra$ are 
\ie\nonumber
1.~&-96 h^3 \left(6+13 h+9 h^2+2 h^3\right)^2 \left(-3-4 h+12 h^2+16 h^3\right)\\
&\times{\left(z^{1+2 h} {\bar{z}}^{2 h} g\left[1-z,1-{\bar{z}}\frac{1}{2},\frac{1}{2},\frac{1}{2}\right]+2 (1-z)^{\frac{1}{2}+2 h} (1-{\bar{z}})^{2
h} g[z,{\bar{z}},0,0,0]\right)}=0\\
2.~&{-24 h \left(6+7 h+2 h^2\right)^2 \left(-3-h+10 h^2+8 h^3\right) }{\left(2 \left(1+3 h+2 h^2\right) z^{1+2 h} {\bar{z}}^{2 h} g\left[1-z,1-{\bar{z}},\frac{1}{2},\frac{1}{2},\frac{1}{2}\right]-\right.}\\
&{h \left((1+h) z^{1+2 h} {\bar{z}}^{2 h} g\left[1-z,1-{\bar{z}},\frac{1}{2},\frac{1}{2},\frac{3}{2}\right]+\right.}{\left.\left.4 (1+2 h) (1-z)^{\frac{1}{2}+2 h} (1-{\bar{z}})^{2 h} g[z,{\bar{z}},0,0,1]\right)\right)}=0\\
3.~&{12 \left(2+5 h+2 h^2\right)^2 \left(-9+28 h^2+16 h^3\right) }{\left((3+2 h) z^{1+2 h} {\bar{z}}^{2 h} g\left[1-z,1-{\bar{z}},\frac{1}{2},\frac{1}{2},\frac{3}{2}\right]-\right.}\\
&{\left.(1+3 h) (1-z)^{\frac{1}{2}+2 h} (1-{\bar{z}})^{2 h} g[z,{\bar{z}},0,0,2]\right)=0}\\
4.~&{8 h^2 \left(3+5 h+2 h^2\right)^2 \left(-3+2 h+8 h^2\right) }{\left(2 \left(2+5 h+2 h^2\right) z^{1+2 h} {\bar{z}}^{2 h} g\left[1-z,1-{\bar{z}},\frac{1}{2},\frac{1}{2},\frac{1}{2}\right]-\right.}\\
&{\left(-1+h^2\right) z^{1+2 h} {\bar{z}}^{2 h} g\left[1-z,1-{\bar{z}},\frac{1}{2},\frac{1}{2},\frac{3}{2}\right]+}{\left.\left.2 (1+2 h)^2 (1-z)^{\frac{1}{2}+2 h} (1-{\bar{z}})^{2 h} g[z,{\bar{z}},0,0,1]\right)\right\} }=0
\fe
The crossing equations obtained from $\la\phi\phi\phi\phi\ra$ are 
\ie
&1.~{z^{2 h} \bar{z}^{2 h} g[1-z,1-\bar{z},0,0,0]-(1-z)^{2 h} (1-\bar{z})^{2 h} g[z,\bar{z},0,0,0]=0}\\
&2.~{z^{2 h} \bar{z}^{2 h} g[1-z,1-\bar{z},0,0,1]-(1-z)^{2 h} (1-\bar{z})^{2 h} g[z,\bar{z},0,0,1]=0}\\
&3.~ {z^{2 h} \bar{z}^{2 h} g[1-z,1-\bar{z},0,0,2]-(1-z)^{2 h} (1-\bar{z})^{2 h} g[z,\bar{z},0,0,2]=0}
\fe
The crossing equations obtained from $\la\psi^2C^2\phi\phi\ra$ are
\ie\nonumber
1.~&{96 h^3 (1+h)^3 \left(-18-9 h+56 h^2+60 h^3+16 h^4\right) }\\
&{\left(2 (1+2 h) z^{2+2 h} \bar{z}^{2 h} g\left[1-z,1-\bar{z},\frac{1}{2},\frac{3}{2},\frac{1}{2}\right]+(2+h) z^{2+2 h} \bar{z}^{2 h}
g\left[1-z,1-\bar{z},\frac{1}{2},\frac{3}{2},\frac{3}{2}\right]+\right.}\\
&{\left.2 (1-z)^{\frac{3}{2}+2 h} (1-\bar{z})^{2 h} ((2+4 h) g[z,\bar{z},-1,0,0]+h g[z,\bar{z},-1,0,1])\right)=0}\\
2.~&{96 h^2 (1+h)^2 \left(-6-11 h+20 h^2+44 h^3+16 h^4\right) }\\
&{\left(2 \left(3+8 h+4 h^2\right) z^{2+2 h} \bar{z}^{2 h} g\left[1-z,1-\bar{z},\frac{1}{2},\frac{3}{2},\frac{1}{2}\right]-\right.}{\left(6+7 h+2 h^2\right) z^{2+2 h} \bar{z}^{2 h} g\left[1-z,1-\bar{z},\frac{1}{2},\frac{3}{2},\frac{3}{2}\right]+}\\
&{\left.(1-z)^{\frac{3}{2}+2 h} (1-\bar{z})^{2 h} \left(4 \left(3+8 h+4 h^2\right) g[z,\bar{z},-1,0,0]-h (2+h) g[z,\bar{z},-1,0,2]\right)\right)=0}
\fe
\ie\nonumber
3.~&{-192 h^2 (1+h)^2 \left(-18-9 h+56 h^2+60 h^3+16 h^4\right) }{\left(2 \left(1+3 h+2 h^2\right) z^{2+2 h} \bar{z}^{2 h} g\left[1-z,1-\bar{z},\frac{1}{2},\frac{3}{2},\frac{1}{2}\right]-\right.}\\
&{h (2+h) z^{2+2 h} \bar{z}^{2 h} g\left[1-z,1-\bar{z},\frac{1}{2},\frac{3}{2},\frac{3}{2}\right]-}{\left.2 (1+2 h) (1-z)^{\frac{3}{2}+2 h} (1-\bar{z})^{2 h} g[z,\bar{z},-1,0,1]\right)=0}\\
4.~&  {48 h \left(-6-17 h+9 h^2+64 h^3+60 h^4+16 h^5\right) }{\left(2 \left(3+11 h+12 h^2+4 h^3\right) z^{2+2 h} \bar{z}^{2 h} g\left[1-z,1-\bar{z},\frac{1}{2},\frac{3}{2},\frac{1}{2}\right]\right.}\\
&{+\left(6+13 h+9 h^2+2 h^3\right) z^{2+2 h} \bar{z}^{2 h} g\left[1-z,1-\bar{z},\frac{1}{2},\frac{3}{2},\frac{3}{2}\right]-}{(1-z)^{\frac{3}{2}+2 h} (1-\bar{z})^{2 h} }\\
&{\left.\left(2 \left(3+11 h+15 h^2+6 h^3\right) g[z,\bar{z},-1,0,1]+h \left(4+8 h+3 h^2\right) g[z,\bar{z},-1,0,2]\right)\right)=0}\\
5.~& {-96 \left(2+5 h+2 h^2\right)^2 \left(-3+2 h+8 h^2\right) }\\
&{\left((3+2 h) z^{2+2 h} \bar{z}^{2 h} g\left[1-z,1-\bar{z},\frac{1}{2},\frac{3}{2},\frac{3}{2}\right]+h (1-z)^{\frac{3}{2}+2 h} (1-\bar{z})^{2
h} g[z,\bar{z},-1,0,2]\right)=0}\\
6.~& {-24 h (1+h)^2 \left(-6-7 h+22 h^2+28 h^3+8 h^4\right) }{\left(-2 h \left(3+10 h+8 h^2\right) z^{2+2 h} \bar{z}^{2 h} g\left[1-z,1-\bar{z},\frac{1}{2},\frac{3}{2},0\right]+\right.}\\
&{\left(-9-33 h+28 h^2+116 h^3+64 h^4\right) z^{2+2 h} \bar{z}^{2 h} g\left[1-z,1-\bar{z},\frac{1}{2},\frac{3}{2},1\right]+}\\
&{\left.2 h (1+2 h)^2 (3+4 h) (1-z)^{\frac{3}{2}+2 h} (1-\bar{z})^{2 h} g\left[z,\bar{z},-1,0,\frac{1}{2}\right]\right)=0}\\
7.~&{-\frac{3}{2} h^2 \left(-2-h+8 h^2+4 h^3\right) }{\left(64 (1+h)^3 \left(9+45 h+22 h^2\right) z^{2+2 h} \bar{z}^{2 h} g\left[1-z,1-\bar{z},\frac{1}{2},\frac{3}{2},1\right]-\right.}\\
&{\left(3+10 h+8 h^2\right) \left(\left(-15+14 h+28 h^2+8 h^3\right) z^{2+2 h} \bar{z}^{2 h} g\left[1-z,1-\bar{z},\frac{1}{2},\frac{3}{2},2\right]+\right.}\\
&{24 (1+h) (1-z)^{\frac{3}{2}+2 h} (1-\bar{z})^{2 h} }{\left.\left.\left(8 (1+h)^2 g\left[z,\bar{z},-1,0,\frac{1}{2}\right]-h (3+2 h) g\left[z,\bar{z},-1,0,\frac{3}{2}\right]\right)\right)\right)=0}\\
8.~& {-12 h (1+h)^2 \left(-2-h+8 h^2+4 h^3\right) }{\left(8 h \left(9+36 h+44 h^2+16 h^3\right) z^{2+2 h} \bar{z}^{2 h} g\left[1-z,1-\bar{z},\frac{1}{2},\frac{3}{2},0\right]-\right.}\\
&{4 \left(-27-117 h-108 h^2+26 h^3+44 h^4\right) z^{2+2 h} \bar{z}^{2 h} g\left[1-z,1-\bar{z},\frac{1}{2},\frac{3}{2},1\right]+}\\
&{h \left(3+10 h+8 h^2\right) (1-z)^{\frac{3}{2}+2 h} (1-\bar{z})^{2 h} }{\left.\left(8 h g\left[z,\bar{z},-1,0,\frac{1}{2}\right]-(3+2 h) g\left[z,\bar{z},-1,0,\frac{3}{2}\right]\right)\right)=0}
\fe
\ie\nonumber
9.~& {3 h^2 \left(2+5 h+2 h^2\right) }{\left(64 (1+h)^3 \left(-9-18 h+32 h^2\right) z^{2+2 h} \bar{z}^{2 h} g\left[1-z,1-\bar{z},\frac{1}{2},\frac{3}{2},1\right]+\right.}\\
&{3 \left(-3-4 h+12 h^2+16 h^3\right) }{\left((5+2 h) z^{2+2 h} \bar{z}^{2 h} g\left[1-z,1-\bar{z},\frac{1}{2},\frac{3}{2},2\right]-\right.}\\
&{\left.\left.8 \left(1+3 h+2 h^2\right) (1-z)^{\frac{3}{2}+2 h} (1-\bar{z})^{2 h} g\left[z,\bar{z},-1,0,\frac{3}{2}\right]\right)\right)=0}
\fe
\end{appendix}
\bibliography{FINALDraft} 
\bibliographystyle{JHEP}

\end{document}